%% file: mol_cloud_sim.tex
\documentclass[twocolumn,twocolappendix]{aastex631}

\usepackage{booktabs}
\usepackage{rotating}
\usepackage{float}
\newcommand{\Rem}{\ensuremath{{\rm Re}_{\rm p}}}
\newcommand{\mG}{\ensuremath{\mu{\rm G}}} % micro Gauss
\newcommand{\turb}{\ensuremath{{\rm turb}}} % Turbulence
\newcommand{\B}{\mathbf{B}}     % Magnetic fields
\newcommand{\E}{\mathbf{E}}     % Electric fields
\newcommand{\J}{\mathbf{J}}   % Current  density
\renewcommand{\v}{\mathbf{v}}   % Velocity fields
\newcommand{\tot}{\mathrm{tot}}
\renewcommand{\O}{\mathrm{O}}     % Ohmic
\newcommand{\A}{\mathrm{A}}     % Ambipolar
\renewcommand{\H}{\mathrm{H}}     % Hall
\renewcommand{\P}{\mathrm{P}}     % P

\input{preamble.tex}
\newcommand{\figdir}{.} 

\begin{document}
 
\title{Turbulent Diffuse Molecular Media with Non-ideal
  Magnetohydrodynamics and Consistent Thermochemistry:
  Numerical Simulations and Dynamic Characteristics}

\author[0000-0003-0355-6875]{Nannan Yue}
\affil{The Kavli Institute for Astronomy
  and Astrophysics, Peking University, Beijing 100871,
  China}
  
\author[0000-0002-6540-7042]{Lile Wang}
\affil{The Kavli Institute for Astronomy and Astrophysics,
  Peking University, Beijing 100871, China}
\affil{Department of Astronomy, School of Physics, Peking
  University, Beijing 100871, China}

\author[0000-0003-2733-4580]{Thomas Bisbas}
\affil{Zhejiang Laboratory, Hangzhou 311121, China}

\author[0000-0003-4811-2581]{Donghui Quan}
\affil{Zhejiang Laboratory, Hangzhou 311121, China}

\author[0000-0003-3010-7661]{Di Li}
\affil{National Astronomical Observatories, Chinese
  Academy of Sciences, Beijing 100101, China} 
\affil{Zhejiang Laboratory, Hangzhou 311121, China}

\correspondingauthor{Lile Wang}
\email{lilew@pku.edu.cn}

\begin{abstract}
  Turbulent diffuse molecular clouds can exhibit complicated
  morphologies caused by the interactions among radiation,
  chemistry, fluids, and fields. We performed full 3D
  simulations for turbulent diffuse molecular interstellar
  media, featuring time-dependent non-equilibrium
  thermochemistry co-evolved with magnetohydrodynamics
  (MHD). Simulation results exhibit the relative abundances
  of key chemical species (e.g., \chem{C}, \chem{CO},
  \chem{OH}) vary by more than one order of magnitude for
  the ``premature'' epoch of chemical evolution
  ($t\lesssim 2\times 10^5~\yr$). Various simulations are
  also conducted to study the impacts of physical
  parameters. Non-ideal MHD effects are essential in shaping
  the behavior of gases, and strong magnetic fields
  ($\sim 10~\mG$) tend to inhibit vigorous compressions and
  thus reduce the fraction of warm gases
  ($T\gtrsim 10^2~\K$).  Thermodynamical and chemical
  conditions of the gas are sensitive to modulation by
  dynamic conditions, especially the energy injection by
  turbulence. Chemical features, including ionization
  (cosmic ray and diffuse interstellar radiation), would not
  directly affect the turbulence power spectra. Nonetheless,
  their effects are prominent in the distribution profiles
  of temperatures and gas densities. Comprehensive
  observations are necessary and useful to eliminate the
  degeneracies of physical parameters and constrain the
  properties of diffuse molecular clouds with confidence.
\end{abstract}


\section{Introduction}
\label{sec:intro}

Molecular clouds are an essential phase of the interstellar
medium (ISM), whose diffuse phase consists mainly of
relatively dense ($\gtrsim 10^2~\cm^{-3}$), cool
($\lesssim 10^2~\K$), and predominantly molecular gases
\citep{DraineBook}. Prominent morphological features,
including filaments, bundles, cloud cores, and occasionally
star-induced bubbles, are shaped by the turbulent fluid
motions interacting with magnetic fields and radiation in
these clouds \citep{2007ARA&A..45..565M,
  2012A&ARv..20...55H}.  Understanding the characteristics
and evolution tracks of the diffuse molecular clouds is
crucial for complete and comprehensive physical pictures of
the evolution of the ISM in general. The diffuse regions
with densities \response{$\rho \lesssim 10^3~m_p~\cm^{-3}$},
which \response{seem} less relevant to subsequent star
formation processes, also draw researchers' attention by
exibiting relatively complicated thermochemical details,
including exitation of ro-vibrational molecular species in
these relatively cold (typically $T \lesssim 10^2~\K$)
regions indicated by existing and incoming observations
\citep[e.g.][]{2008A&A...483..471N, 2010ApJ...715.1370G,
  2010A&A...521L..17L, 2011ApJ...743..174I,
  2018ApJ...858..111R}.

% and the subsequent ``dark clouds'' stage for star
% formation processes in particular.

Multiple numerical studies have been conducted regarding
interstellar molecular gas's morphological and dynamical
characteristics. Pioneered by, e.g.,
\citet{1998PhRvL..80.2754M, 1999ApJ...524..169m}, a series
of studies on molecular clouds using 3D hydrodynamic or
magnetohydrodynamic (MHD) simulations have focused on the
morphologies and statistical characteristics of decaying or
quasi-steady turbulence structures. Plenty of simulation
works have assumed isothermal equations of states, even some
of them have adopted ray-tracing or diffusive radiative
transfer \citep{2016ApJ...829..130R, 2017ApJ...850..112R,
  2018ApJ...859...68K, 2019ApJ...883..102K}.  Nonetheless,
consistent cooling and heating processes are considered
essential to the reaction rates and excitation states of
molecules, as well as the dissipation conditions for the
turbulence scaling laws \citep{1995ApJS..100..132N,
  2007ApJ...660.1332J, 2007ApJ...666....1G,
  2011MNRAS.415.3681T, 2018ApJ...864..116Q,
  2018A&A...611A..20W}. Simulating the turbulent media with
such global prescriptions, rather than real-time local
calculations, risks obscuring the interactions between
thermochemistry and dynamics, leading to potentially
unreliable consequences in dynamics and observational
predictions.

Another critical factor is the non-ideal MHD effects. The
weak ionization in molecular clouds leads to finite
tensorial conductivities, exhibiting deviations from the
flux-freezing ideal MHD. The tensor field of these magnetic
diffusivities depends on the spatial distributions of plasma
temperatures, free electrons, positive ions, and charged
dust particles. Because of the lack of detailed
thermochemical information, most simulation works tend to
prescribe, rather than calculate in real-time, a set of
conductivity or diffusivity parameters
\citep{2009ApJ...701.1258D, 2011MNRAS.415.3681T,
  2011MNRAS.418..390J, 2012MNRAS.420..817J}. Even in the
star formation simulations, magnetic diffusivities play
crucial roles, \citet{2016MNRAS.457.1037W} adopted
significantly simplified equilibrium chemistry for charge
carriers and ignored thermodynamics. However, the
complications in charge-related reactions (especially those
involving molecules and dust grains) and radiation cooling
still necessitate real-time non-equilibrium thermochemistry
despite the high computational costs, as have been
identified in multiple studies on protoplanetary disks and
molecular clouds \citep{2016ApJ...819...68X,
  2019ApJ...874...90W, 2019MNRAS.486.4622C}.

In order to address these critical issues for better
intrinsic consistency and predictability for observations,
we adopt the GPU-accelerated co-evolving non-ideal
MHD-thermochemistry system that has been used in simulating
protoplanetary disks \citep[e.g.][]{2019ApJ...874...90W} and
predicting exoplanetary atmosphere observations
\citep{2021ApJ...914...98W}. There are several recent
simulation studies that are conceptually similar.  Works by
e.g. \citet{2019MNRAS.489.1719W, 2024MNRAS.528.2257W} have
delved into the use of magnetic diffusivity profiles derived
from chemical equilibrium to investigate the formation of
stellar cores and protostellar disks in relatively dense
molecular environments. Their findings underscore the
critical role of magnetic diffusivities in addressing the
magnetic braking issue. Other studies, such as
\citet{2018MNRAS.481.4277F, 2019MNRAS.486.4622C,
  2021A&A...654L...6L}, have examined various facets of
molecular gases through the lens of non-equilibrium
chemistry, focusing on the evolution of observables and
tracer species. However, these studies have not taken into
account the non-ideal MHD effects.  Our approach aims to
bridge this gap by conducting a comprehensive series of
calculations within a dynamically quasi-steady turbulent
framework. We will employ consistent magnetic diffusivities
that are grounded in more extensive, non-equilibrium
thermochemical networks in realtime. This endeavor aims at
providing a more thorough exploration covering the lifecycle
of typical diffuse molecular ISM, which spans approximately
$\gtrsim 10^6~\yr$. Our goal is to reflect the long-term
evolution of thermochemical processes, thereby offering a
more nuanced and accurate representation of the ISM's
behavior over extended periods. We will portray diffuse
molecular clouds from diverse physical perspectives and
systematically explore the space spanned by key physical
parameters, especially the interstellar radiation field
(ISRF), the cosmic rate (CR) ionization rate, the mean-field
magnetic intensity, the specific dust cross-section,
metallicity, and the turbulence energy injection rate. With
detailed real-time thermochemical networks evolved and
recorded, extended explorations on the observables and
characterizations can also stem from these simulations in
subsequent future works.

This paper is structured as
follows. Section~\ref{sec:methods} briefly introduces the
methods used in the simulations and describes the setups of
several representative simulations. Setups and results of
the fiducial simulation are elaborated in
Section~\ref{sec:fiducial}, and various models over the
physical parameter space are explored in 
Section~\ref{sec:var-model}. We discuss the possible future
explorations based on current results and conclude the paper
in Section~\ref{sec:summary}. 

\section{Methods and Setups}
\label{sec:methods}

The computational schemes summarized in this section are
primarily similar to \citet{2019ApJ...874...90W}, with
modifications particular to molecular clouds. We will also
discuss the setups of numerical models in this paper.

\subsection{Non-ideal Magnetohydrodynamics}
\label{sec:method-mhd}

The ionization of interstellar media, including the diffuse
molecular gas explored in this paper, is always
significantly non-zero. This fact makes their morphologies,
evolution tracks, and thermochemistry profiles susceptible
to different interactions between fluids and fields. We use
the grid-based higher-order Godunov MHD code \verb|Athena++|
\citep{2020ApJS..249....4S} on uniform Cartesian grids, with
HLLD approximate Riemann solver and piecewise linear method
(PLM) reconstruction. The MHD equations are solved in the
conservative form:
\begin{equation}
  \label{eq:mhd}
  \begin{split}
    & \partial_t \rho + \nabla \cdot (\rho \v) = 0\ ;
    \\
    & \partial_t (\rho \v) + \nabla \cdot \left(\rho \v \v -
      \dfrac{\B\B}{4\pi} + P_\tot \mathbf{I} \right) = -
    \nabla \Phi \ ;
    \\
    & \partial_t \B = \nabla \times (\v \times \B - c \E' )\
    ;
    \\
    & \partial_t \epsilon + \nabla \cdot \left[
      \left(\epsilon + P_\tot \right) \v -
      \dfrac{(\B\cdot\v)\B}{4\pi} \right]
    = S\ ,
  \end{split}
\end{equation}
where $\rho$, $\v$ and $p$ are the gas density, velocity and
gas thermal pressure respectively, $\B$ is the magnetic
field, $P_\tot\equiv p + B^2/(8\pi)$ is the total pressure,
$\Phi$ is the gravitational potential, $\mathbf{I}$ is the
identity tensor, and
\begin{equation}
  \epsilon \equiv \dfrac{p}{\gamma - 1} + \rho
  \left( \dfrac{v^2}{2} + \Phi \right) + \dfrac{B^2}{8\pi}
\end{equation}
is the total energy density. The extra source term $S$ in
the energy equation reflects non-adiabatic processes that
affect gas energy, which is calculated separately from the
MHD solver in an operator-splitting manner (see
\S\ref{sec:method-micro}). The adiabatic index $\gamma$ is
evaluated based on the real-time chemical components in each
simulation zone.  Non-ideal MHD effects are also considered
by including an electric field in the local rest frame of
fluids,
\begin{equation}
  \label{eq:non-ideal-def}
  \E' = \dfrac{4\pi}{c^2}
  \left( \eta_\O\J + \eta_\H \J\times \mathbf{b} + \eta_\A
    \J_\bot \right)\ ,
\end{equation}
where $\mathbf{b}\equiv \B/B$ is a unit vector along the
local direction of $\B$, $\J = c \nabla \times \B / 4\pi$ is
the current density, and
$\J_\bot\equiv \mathbf{b}\times (\J \times \mathbf{b})$ is
the component of $\J$ perpendicular to the local magnetic
field. Different diffusivity components, including Ohmic
($\eta_\O$), ambipolar ($\eta_\A$), and Hall ($\eta_\H$),
are calculated for each simulated zone with the
thermochemical conditions also elaborated in
\S\ref{sec:method-micro}.

\subsection{Non-equilibrium Thermochemistry with Consistency}
\label{sec:method-micro}

The thermochemical processes significantly modulate the
dynamic properties in turbulent molecular zones, which are
also important in observational measurements and
diagnostics. Technically, a ``flow reaction'' paradigm is
required, in which the MHD is in charge of advecting
chemical species and manipulating internal energy around the
simulated space. Thermochemical reaction networks are
co-evolved with the MHD in every simulation zone.

The advection of chemicals with MHD is accomplished by
specific methods so that consistency of chemical elements and
species fluxes in every timestep of MHD evolution is
ensured. The advected chemical abundances and internal
energy are then lead to the thermochemical solver for the
current step, solving the following coupled set of ordinary
differential equations (ODEs)  (in which the
Einstein summation convention applies):
\begin{equation}
  \label{eq:ode-chem-thermo}
  \dfrac{\d n^i}{\d t} = \mathcal{A}^i_{\;jk} n^j n^k +
  \mathcal{B}^i_{\;j} n^j\ ; \quad
  \dfrac{\d \epsilon}{\d t} = \Gamma - \Lambda\ ;
\end{equation}
in which the terms $\{\mathcal{A}^i_{\;jk}\}$ describe
two-body reactions, those in $\{\mathcal{B}^i_{\;j}\}$
represent photoionization, photodissociation, and
spontaneous decays, and $\Gamma$ and $\Lambda$ represent all
non-adiabatic heating, cooling, and heat transfer rates per
unit volume, respectively. We use a multi-step implicit
method with error control to solve since the ODEs
\eqref{eq:ode-chem-thermo} are usually stiff. The costs of
co-evolving a stiff network in each zone throughout the
domain at every timestep could be excessively high. For
instance, a 3D simulation domain with $128^3$ zones has to
co-evolve $\sim 2\times 10^6$ stiff systems in each step.
In order to finish the simulations within reasonable
``wall-clock time'', we use the thermochemistry solver
implemented on graphic processing units (GPUs) to accelerate
the simulations considerably for reasonable computing costs
(see also \citealt{2017ApJ...847...11w,
  2019ApJ...874...90W}).

The set of chemical species, reactions, and thermodynamic
processes largely follows \citet{2019ApJ...874...90W}, with
updates specific to diffuse molecular zones. The 23 species
are: \neg{e} (free electrons), \chem{H^+}, H, \chem{H_2},
\chem{H_2^+}, He, \chem{He^+}, O, \chem{O^+}, OH,
\chem{H_2O}, \chem{H_2O^+}, \chem{H_3O^+}, C, \chem{C^+},
CO, \chem{HCO^+}, \chem{CH}, \chem{CH_2^+}, Gr, \chem{Gr^+},
\chem{Gr^-}. Here Gr and Gr$^\pm$ denote neutral and
singly-charged dust grains, respectively. This set covers
the most concerned chemical species in astronomical
measurements. Internal energy density is also added as the
last ``species'' to the set, and the temperature dependences
of $\{\mathcal{A}^i_{\;jk}\}$ and $\{\mathcal{B}^i_{\;j}\}$
coefficients are converted to internal energy dependences
(via proper heat capacities) for consistent thermochemical
co-evolution.  These ``species'' interact with each other
through a network with 146 chemical reactions and
thermodynamic processes (cooling and heating), most of which
are included in the \verb|UMIST| database
(\citealt{umist2013}), and other necessary processes are
also involved. To name a few mechanisms that are potentially
important to the diffuse molecular cloud evolution, we have:
\begin{itemize}
\item Photoionizations by the interstellar radiation field
  (ISRF) and chemical species, including the photoionization
  of \chem{C}, and photoelectric effects of dust grains.
\item The ISRF photodissociation of multiple molecular
  species, e.g., \chem{OH}, \chem{H_2O}, \chem{CO}, and
  \chem{H_2}.
\item Cosmic ray (CR) ionization of multiple neutral
  species, especially \chem{H_2}, \chem{H}, \chem{He}, and
  \chem{O}. 
\item Formation of important tracer molecules such as
  \chem{CO} and \chem{OH}, as subsequent products of ISRF
  photoionized \chem{C^+}, and CR ionized \chem{O^+} (see
  also \citealt{DraineBook}).
\item Heating processes, including the energy injected via
  ISRF photoionization and dissociation, CR ionization, and
  the \chem{H_2} formation on the surfaces of dust grains
  ($\sim 1~\eV$ per molecule formed).
\item Radiative cooling processes, including ro-vibrational
  transitions of \chem{CO}, \chem{OH}, and \chem{H_2}
  \citep{1993ApJ...418..263N, 2010ApJ...722.1793O}, and the
  fine structure transition of \chem{C^+}.
\end{itemize}
Note that the self-shielding effects for relevant
photoionization and photodissociation processes are included
with the recipes in \citet{2017A&A...602A.105H}.

Both the radiation-related reactions and radiative cooling
processes involve the penetration of photons, and column
densities of different chemical species for photons to
impinge and escape are the most important physical
parameter. For the local structure simulations using
periodic boundary conditions, the penetration columns for
photons cannot be obtained consistently. Therefore, we
estimate the column densities of species X involved in
shielding is estimated by
$N(\chem{X}) \sim n(\chem{X}) \bar{l}_v$, and the effective
length $l_v$ is approximated by
\begin{equation}
  \label{eq:equiv-length}
  \bar{l}_v \sim \min \left\{ l_v,\ l_{\rm cap} \right\}\ ;
  \quad l_v \equiv \dfrac{|\mathbf{v}|}
  {|\nabla\cdot\mathbf{v}|}\ ,
\end{equation}
where $l_{\rm cap}$ approximates the distance to the nearest
source of radiation, describing case that the gas is largely
uniform and static. There are also alternatives to this
Sobolev extimation of effective length, e.g.,
$l_\rho\equiv \rho/|\nabla \rho|$ and
$\bar{l}_\rho\sim \min\{l_\rho,\ l_{\rm cap}\}$. However,
the estimates based on velocity best emulates the physical
mechanisms involved for the penetration of cooling and
ionizing photons. For the diffuse region in the Taurus
molecular cloud (TMC hereafter), we assume that
$l_{\rm cap}\sim 1~\pc$ represents the typical distance to
one of the related radiation sources (e.g. young stars) in
the nearby regions. In addition, the typical sizes of clumps
inside the TMC diffuse regions are also considerably smaller
than $\sim 1~\pc$ \citep[see also][]{2010ApJ...715.1370G},
making $l_{\rm cap}\sim 1~\pc$ a resonably capping for the
photon penetration length. We will revisit and confirm this
assumption when analyzing the simulation results.

Similar to \citet{2019ApJ...874...90W}, non-ideal MHD
diffusivities are determined by the gas temperature and the
charge carriers' abundances computed via the real-time
non-equilibrium thermochemical network.  The general
expressions for the three diffusivities read
\citep[e.g.][]{2011ApJ...739...51B},
\begin{equation}
  \label{eq:def-diffusivity}
  \begin{split}
    & \eta_\O =
    \dfrac{c^2}{4\pi}\left(\dfrac{1}{\sigma_\O}\right)\
      ,\quad 
    \eta_\H =
    \dfrac{c^2}{4\pi}
    \left(\dfrac{\sigma_\H}{\sigma_\H^2+\sigma_\P^2}\right)\,
    \quad 
    \\
    & \eta_\A =
    \dfrac{c^2}{4\pi}
    \left(\dfrac{\sigma_\P}{\sigma_\H^2+\sigma_\P^2}\right)
    - \eta_\O\ ,
  \end{split}
\end{equation}
where $\sigma_\O$, $\sigma_\H$ and $\sigma_\P$ are Ohmic,
Hall, and Pederson components of conductivity under magnetic
fields, and
\begin{equation}
  \label{eq:def-conductivity}
  \begin{split}
    & \sigma_\O = \dfrac{ec}{B} \sum_j n_jZ_j\beta_j
    \ ,\quad
    \sigma_\H = \dfrac{ec}{B} \sum_j
    \dfrac{n_jZ_j}{1+\beta_j^2}\ ,
    \\
    & \sigma_\P = \dfrac{ec}{B} \sum_j
    \dfrac{n_jZ_j\beta_j}{1+\beta_j^2} ,
  \end{split}
\end{equation}
in which $Z_je$ is the charge and $n_j$ the number density
of the $j^{\rm th}$ charged species, and the Hall parameter
$\beta_j$ is the ratio of the gyrofrequency to the collision
rate with neutrals,
\begin{equation}
  \label{eq:def-hall-param}
  \beta_j = \dfrac{Z_jeB}{m_jc}
  \dfrac{1}{\gamma_j\rho}\ ;\quad \gamma_j \equiv
  \dfrac{\mean{\sigma v}_j}{\mean{m}_n + m_j}\ ,
\end{equation}
where $m_j$ is the charged species' molecular mass,
$\mean{m}_n$ is the mean molecular mass of the neutrals, and
$\mean{\sigma v}_j$ is the rate of collisional momentum
transfer between the $j$th species and the neutrals.

\subsection{Turbulence Driving}
\label{sec:method-turb}

From larger scales, turbulence energy cascades down to the
simulated systems--a localized, periodic, cubic domain--in
this work. Various methods have been adopted to emulate the
injection of turbulence energy, and we adopt the mature
methods adopted by \citet{1999ApJ...524..169m}. The
perturbations are inserted as a globally uniform
acceleration or inertial force. At each time step, a unity
vector with random direction $\hat{a}$ is assigned
throughout the domain for the direction of
perturbations. Assuming the amplitude of perturbation is
$A$, the turbulence energy injection rate per by unit mass
$\dot{\epsilon}_\turb$ can be evaluated as,
\begin{equation}
  \label{eq:turb-inj-norm}
  \dot{\epsilon}_\turb = A \left\{ [\mean{\rho} L_{\rm box}^3]^{-1}
  \Delta t \  \sum_{\mathbf{i}} \delta V_{\mathbf{i}}\;
  \rho_{\mathbf{i}} \mathbf{v}_\mathbf{i} \cdot \hat{a}\right\}\ ,
\end{equation}
where $L_{\rm box}$ is the box length, $\mean{\rho}$ is the
mean mass density throughout the domain, $\Delta t$ is the
current timestep, $\mathbf{i}$ is the cell index,
$\delta V_\mathbf{i}$, $\rho_\mathbf{i}$ and
$\mathbf{v}_\mathbf{i}$ are the cell volume, mass density,
and velocity vector for the $\mathbf{i}$-th cell, and the
summation goes through all cells. In practice, we specify a
fixed $\dot{\epsilon}_\turb$ as one of the model parameters
throughout the evolution period. In each time step, we use
the summation in the braces of eq.~\eqref{eq:turb-inj-norm}
to determine the normalization factor $A$, calculate the
perturbation acceleration vector $A\hat{a}$, inject kinetic
energy into the system by
$\mathbf{v}_\mathbf{i}' = \mathbf{v}_\mathbf{i} + A\hat{a}
\Delta t$, and adjust the total energy density accordingly.

\subsection{Simulation Setups}
\label{sec:method-setup}

\begin{deluxetable}{lr}
  \tablecolumns{2}  
  \tabletypesize{\scriptsize}
  \tablewidth{0pt}
  \tablecaption{Properties of the Fiducial Model
    (\S\ref{sec:method-setup}) 
    \label{table:fiducial-model}
  } \tablehead{ \colhead{Item} & \colhead{Value} }
  \startdata
  Geometries & \\
  Simulation domain & $(0.04~{\rm pc})^3$\\
  Resolution & $N^3 = 128^3$ \\
  \\
  Irradiation & \\
  Interstellar radiation field (ISRF) ${}^\dagger$ & $0.3\;
  G_0$ \\
  Cosmic ray ionization ${}^*$ & $\zeta = 10^{-17}~\s^{-1}$ \\
  \\
  Initial MHD Parameters & \\
  Temperature & $30~\K$ \\
  Density & $250~m_p~\cm^{-3}$ \\
  Magnetic Field & $3~\mG$ \\    
  \\
  Initial abundances [$n_{\chem{X}}/n_{\chem{H}}$] & \\
  \chem{H_2} & 0.5\\
  \chem{He} & 0.1\\
  \chem{H} & $10^{-3}$\\
  $e^-$ & $10^{-6}$\\
  \chem{H^+} & $10^{-6}$\\  
  \chem{O} & $3.2 \times 10^{-4}$\\
  \chem{C} & $1.4 \times 10^{-4}$\\
  \\
  Dust/PAH properties & \\
  $a_\Gr$ & $5$ \AA \\
  $\sigma_\Gr/\chem{H}$ & $\sim 10^{-23}~\cm^2/\H$ \\
  \enddata
  \tablecomments{$\dagger$: $G_0$ denotes the Habing
    radiation flux (see also \citealt{DraineBook}).\\
    $*$: Total ionization rate per hydrogen atom by cosmic
    ray; the branching ratio of ionization is proportional
    to the elemental abundances (see
    e.g. \cite{bai+goodman2009}).  }
\end{deluxetable}

In brief, we carry out 3D MHD non-equilibrium thermochemical
simulations in a (0.04 pc)$^3$ box with periodic boundary
conditions.  Non-equilibrium thermochemistry, whose network
has 23 chemical species (plus internal energy density) and
146 reactions, is co-evolved in every zone of the MHD
simulation. Non-ideal MHD effects with all components of
tensorial conductivities are included.

For the diffuse region outside the boundary in the TMC, the
ISRF UV photodissociation and photoionization parameter is
$0.3~G_0$ \citep {2010A&A...521L..19P}.  At $t=0$, we set
the initial uniform mass density to be
$\rho_{t=0} = 250~m_p~{\rm cm}^{-3}$ with $m_p$ the proton
mass, and the initial temperature to be uniformly
$T_{t=0} = 10~{\rm K}$ throughout the entire simulation
domain. We also put a $\mathbf{B}_{t=0}=B_0 \mathbf{e}_x$
initial magnetic field, in which $B_0 = 3~\mu{\rm G}$ for
this fiducial model. We let the system evolve without
additional energy input for $10^6~{\rm yr}$ and examine the
decay of the initial velocity perturbations and the gas
cooling to select a time that best represents the
thermochemical state of the TMC. These properties define the
fiducial model, summarized in Table~\ref
{table:fiducial-model}. Using these properties on
hydrostatic models, we obtain two test cases that verify the
applicability of our simulation system on molecular
clouds. These tests, which also serve as the reference point
of our subsequent analyses of thermochemistry for models
{\it with} turbulences, are elaborated in
Appendix~\ref{sec:method-static-verify}.

\section{Fiducial Model Results}
\label{sec:fiducial}

\begin{figure}
  \centering
  \hspace*{-0.4cm}
  \includegraphics[width=\linewidth]
  {\figdir/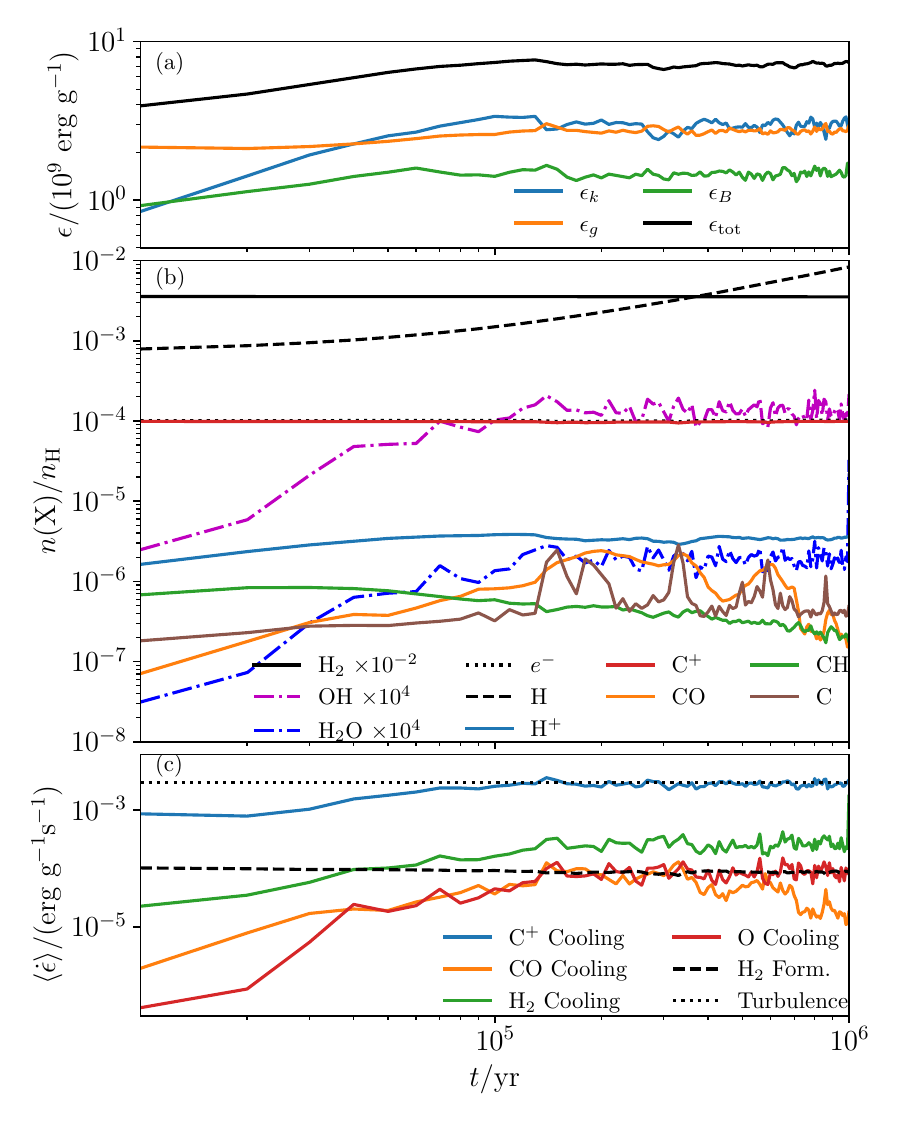}
  \caption{Time evolution of key physical parameters of the
    fiducial model. {\bf Upper panel} (a): Domain-averaged
    specific energies (energy per unit mass; $\epsilon_k$
    for the kinetic energy, $\epsilon_g$ for the gas
    internal energy, $\epsilon_B$ for magnetic energy, and
    $\epsilon_{\rm tot}$ for the sum of all three
    components). {\bf Middle panel} (b): Relative abundances
    of a few key chemical species throughout the simulation
    domain (note that the abundances of \chem{OH} and
    \chem{H_2O} are multiplied by $10^4$). {\bf Lower panel}
    (c): Average energy budgets per unit mass (solid lines
    for cooling, dash and dotted lines for heating or
    injections). }
  \label{fig:history_fiducial}
\end{figure}

\begin{figure}
  \centering
  \hspace*{-0.4cm}
  \includegraphics[width=3.6in]
  {\figdir/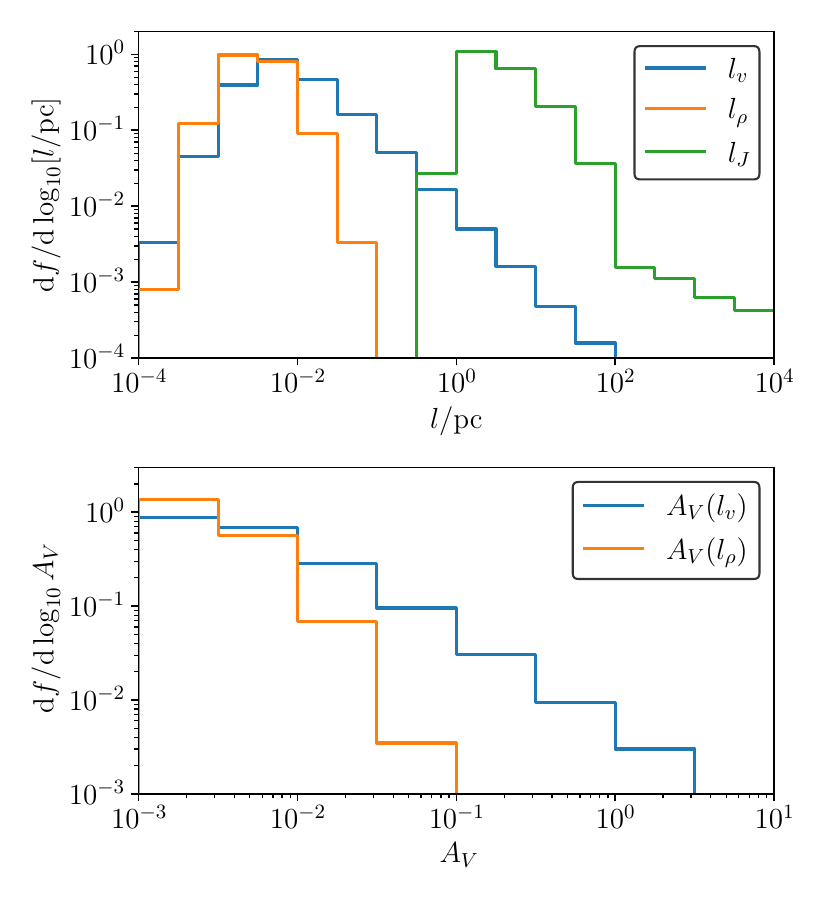}
  \caption{{\bf Upper panel}: Distribution of volume
    fractions about the logarithm of scale lengths
    ($\d f/\d\log_{10} l$), including extinction scale
    lengths estimated by velocity divergence ($l_v$),
    density gradient ($l_\rho$), and the Jeans length
    ($l_J$). {\bf Lower panel}: Distribution of volume
    fractions about the logarithm of visual extinction,
    estimated by \response{
      $A_V\sim N_\H/(1.8\times 10^{21}~\cm^{-2})$} using
    $l_v$ and $l_\rho$ for the effective column densities,
    respectively. }
  \label{fig:column_fiducial}
\end{figure}

The evolution history of the fiducial model (Model 0) is
illustrated in Figure~\ref{fig:history_fiducial}. The
dynamical evolution, characterized by different components
(internal energy $\epsilon_g$, kinetic energy $\epsilon_k$,
and magnetic energy $\epsilon_B$) of the specific energy,
has mostly reached its quasi-steady state since
$t\sim 10^5~\yr$. After that, the energy balance only
increases at a shallow rate ($\sim 10\%$ over
$\sim 10^6~\yr$). The thermochemistry evolution,
nonetheless, has longer timescales. As the most important
coolants for material in the diffuse molecular phase,
variations in the abundances of a few key chemical species
(e.g., \chem{CO}, \chem{C}, \chem{OH}) fundamentally
modulate the cooling channels in the thermodynamic budgets
of the simulated gas. Overall, the comparison in timescales
emphasizes the necessity of using non-equilibrium
thermochemistry. In Figure~\ref{fig:column_fiducial} we
confirm that the effective penetration length estimated by
velocity divergences and density gradients yield results
that are qualitatively consistent, which fall well below the
capping length $l_{\rm cap}$. The visual extinction is
mostly negligible ($A_V < 0.5$ or even $\lesssim 0.1$) for
$\gtrsim 99.5\%$ of the simlated zones, which is consistent
with the studies for the diffuse regions TMC
\citep[e.g.][]{2010ApJ...715.1370G}. In addition, the Jeans
length is significantly greater than the box scales even for
compressed regions in the fiducial model. In addition, the
virial parameter
\begin{equation}
  \alpha_{\rm vir}\sim \dfrac{ v^2+c_s^2}{G\rho
    l^2}\ ,
\end{equation}
are also large within the simulation domain. The median of
$\alpha_{\rm vir}$ reads
$\langle\alpha_{\rm vir}\rangle_{\rm median}\sim 3$ with
$l\sim l_J$ over the Jeans length, and
$\langle\alpha_{\rm vir}\rangle_{\rm median}\sim 10^6$ over
the typical clump scales estimated with $l\sim l_\rho$.
These values indicate that the simulated diffuse regions are
gravitationally unbound, and neglecting the self-gravity is
eligible for the analyses on dynamics in this paper.

\subsection{Fluid and Field Properties}
\label{sec:fiducial-mhd}

\begin{figure*}
  \centering
  \includegraphics[width=2.52in, keepaspectratio]
  {\figdir/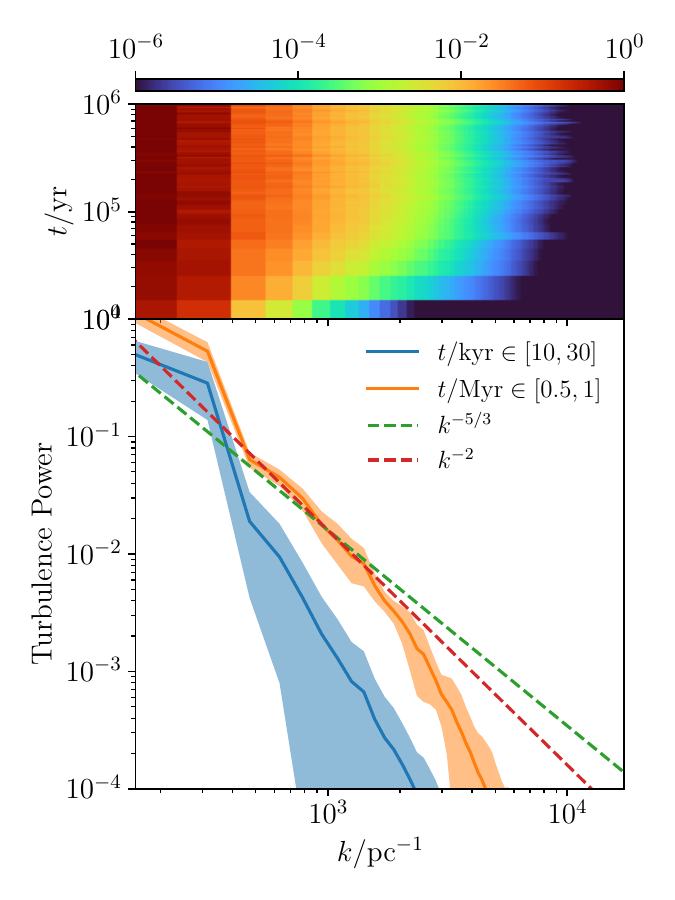}
  \hspace{-0.8cm}
  \includegraphics[width=2.52in, keepaspectratio]
  {\figdir/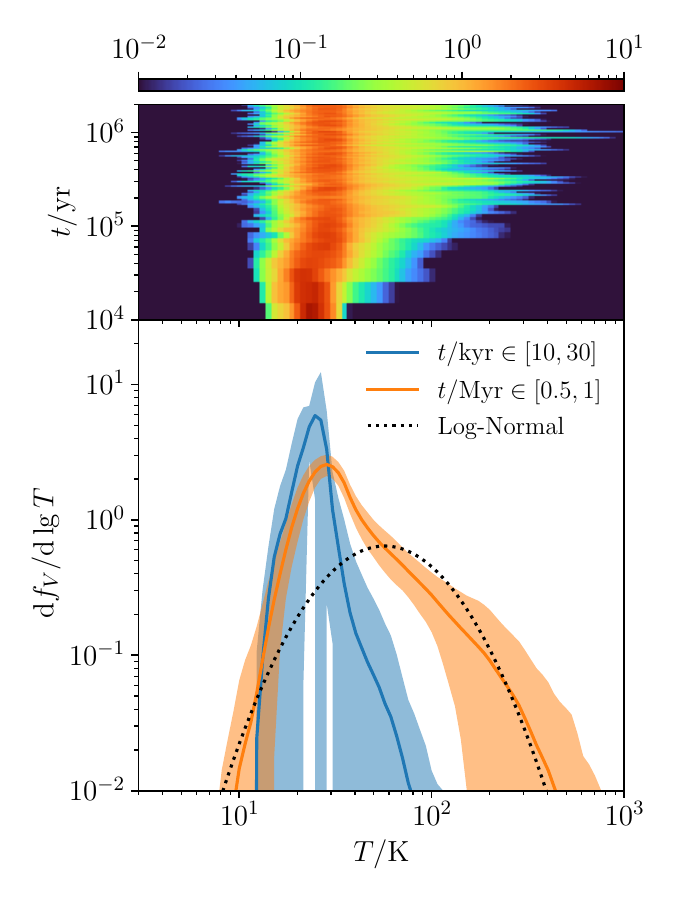}
  \hspace{-0.8cm}  
  \includegraphics[width=2.52in, keepaspectratio]
  {\figdir/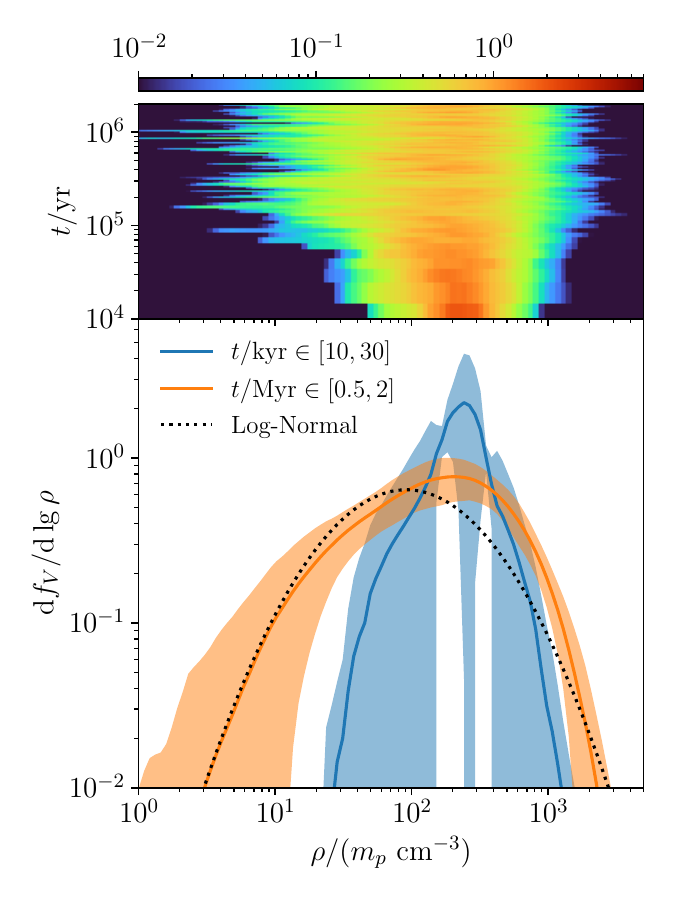}  
  \caption{Turbulence kinetic energy power spectra for
    fiducial model. {\bf Upper row}: Evolution of statistics
    over time; {\bf Lower row}: Time-averaged statistics in
    different evolution epochs; shaded bands illustrate the
    ranges of $2\sigma$ variation. The left column shows the
    power spectra (whose lower panel compares them to the
    standard Kolmogorov $k^{-5/3}$ power law, and the
    $k^{-2}$ power law), the middle column shows the
    distributions of spatial volume fraction $f_V$ in the
    $\lg T$ space, and the right column shows the
    distributions in the $\lg \rho$ space. Dotted curves on
    the lower-middle and lower-right panels present the
    typical shapes of log-normal distributions for
    references.  }
  \label{fig:evo_fiducial}
\end{figure*} 

\begin{figure*}
  \centering
  \includegraphics[width=7.2in, keepaspectratio]
  {\figdir/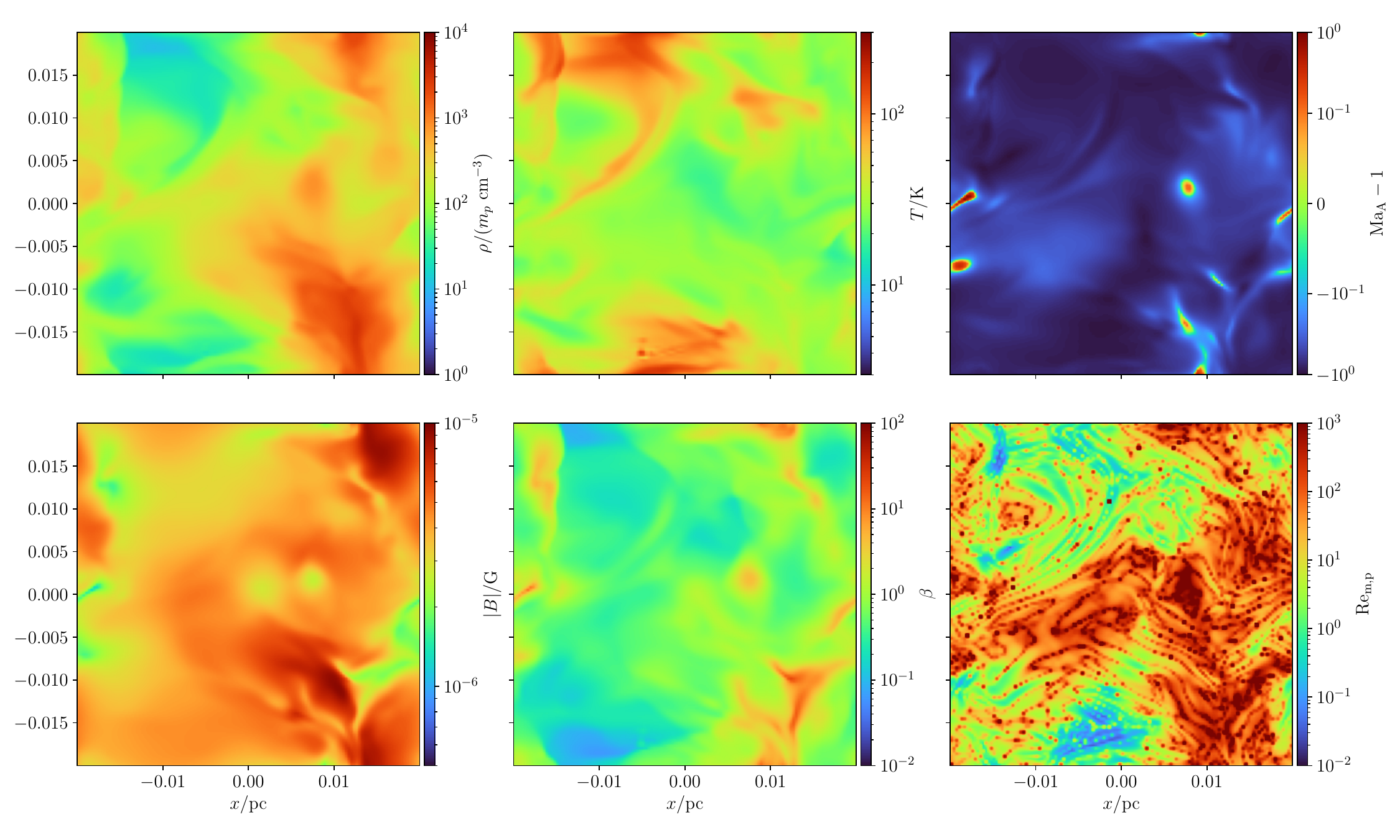}
  \caption{Slices of physical parameters at the $z=0$ plane,
    illustrating the $t=10^6~\yr$ snapshot for the fiducial
    model. The upper row shows the key hydrodynamical
    parameters (mass density $\rho$, temperature $T$, and
    ${\rm Ma}_\A - 1$
    (${\rm Ma}_\A\equiv |v|/[B/(4\pi\rho)^{1/2}]$ is the
    Alfv\'enic Mach number), and the lower row shows
    relevant field parameters (magnetic field strength
    $|B|$, plasma $\beta$ parameter, and magnetic Reynolds
    number $\Rem$). }
  \label{fig:slice_fiducial}
\end{figure*}

\begin{sidewaysfigure*}
  \centering
  \vspace{8cm}
  % \vspace{-10cm}  
  \includegraphics[width=9.2in, keepaspectratio]
  {\figdir/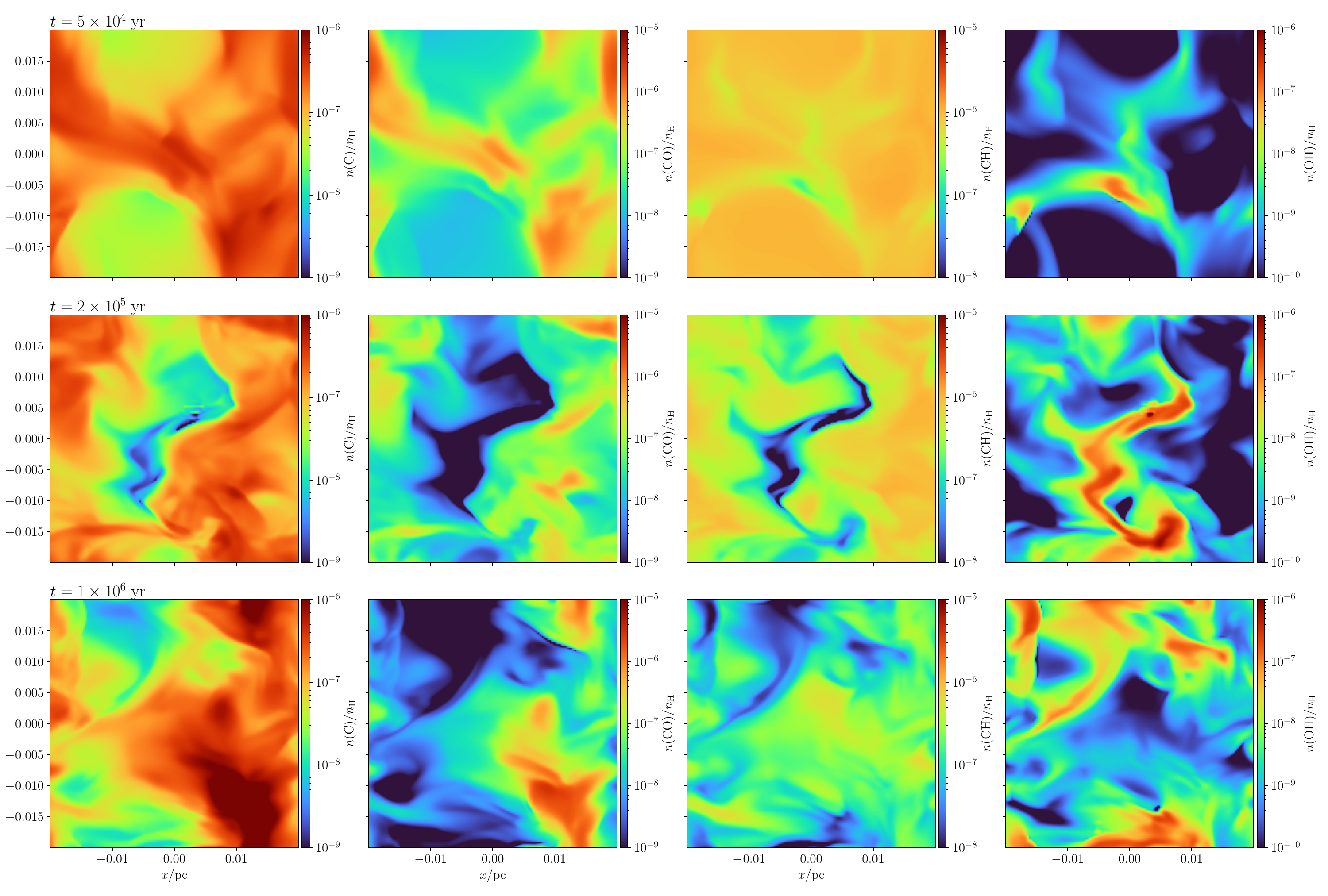}
  \caption{Slices at the $z=0$ plane for the fiducial for
    different evolution time ($t=5\times 10^4~\yr$ for the
    upper row, and $5\times 10^5~\yr$ for the lower row),
    showing the relative abundances of \chem{C} (left
    column), \chem{CO} (middle-left column), \chem{CH}
    (middle-right column), and \chem{OH} (right
    column), normalized to the total density of hydrogen
    nuclei $n_\H$. }
  \label{fig:chem_fiducial}
\end{sidewaysfigure*}

As the initial fields was set uniformly $B_{x0} = 3~\mG$,
the corresponding initial plasma $\beta$ was
$\beta_0 = 1.23$. Turbulent evolution injects kinetic energy
directly into the box, which subsequently branches into
thermal energy by shocks, viscous dissipations, and magnetic
energy via flux freezing. Similar to
Figure~\ref{fig:history_fiducial}, the evolution of kinetic
energy spectrum and distribution functions (for $\lg T$ and
$\lg \rho$) illustrated in Figure~\ref{fig:evo_fiducial}
reaches quasi-steady states after $t\gtrsim 10^5~\yr$.

The kinetic energy spectrum does not show significant
variation since $t\gtrsim 10^5~\yr$, confirming that the
quasi-steady state has already been reached. Magnetization
for the fiducial model is intermediate ($\beta\sim 1$), with
most of the gas dynamics being sub-Alfv\'enic
(Figure~\ref{fig:slice_fiducial}) but supersonic (not shown
in the figure). We observe a power law function damped at
$k\sim 1.3\times 10^3~\pc^{-1}$ (or equivalently
$\sim 0.005~\pc$ spatial scale). At lower wavenumbers down
to $k\sim 0.4\times 10^3~\pc^{-1}$ (lager scales at
$\sim 1.5\times 10^{-2}~\pc$) the spetrum fits the
Kolmogorov $E(k)\propto k^{-5/3}$ law. On a wider range of
wavenumbers spanning roughly one decade (up to a damping
scale $k\sim 1.5~\times 10^3~\pc^{-1}$), the energy power
spectrum is comparable to the $E(k)\propto k^{-2}$ Burgers
power law (for weakly magnetized compressible plasma with
shocks, e.g., \citealt{2007ApJ...658..423K,
  2009PhRvL.103v5001B}).  The fitting over the
$k/(10^3~\pc)\in [0.5, 1.5]$ range of wavenumbers prefers
the $k^{-2}$ power index ($\chi^2/N\sim 0.7$), and varying
the wavenumber range included yields power index fittings
roughly ranging from $-1.9$ through $-2.1$. Compared to the
fluid morphologies in the configuration space
(Figure~\ref{fig:slice_fiducial}), the damping scale is
close to the sizes of a typical eddy ($\sim
0.003~\pc$). These sizes are significantly greater than the
numerical dissipation scale (roughly a few times the cell
size, $\Delta x \simeq 3\times 10^{-4}~\pc$).  Therefore,
the dissipation should be physical rather than numerical or
artificial. \S\ref{sec:var-model-mag} re-examines such
dissipation scales from the aspect of fields.

The distribution functions of $\lg T$ and $\lg\rho$ exhibit
more significant variations than the power spectrum. The
upper branch of the $\lg T$ distribution and the lower
branch of the $\lg\rho$ distribution mostly fit the
log-normal distributions commonly observed in
turbulences. In contrast, the other sides (low temperatures
and high densities) consist of the peaks above the
log-normal components that result directly from cooling
processes. We notice $\gtrsim 10\%$ of the simulated domain
(volume fraction) whose temperature is warm, $T>
10^2~\K$. From the slices of the fiducial model in
Figure~\ref{fig:slice_fiducial}, we notice that these
relatively warm regions usually have lower (but not
necessarily lowest) mass densities. Such temperatures will
likely populate the ro-vibrational energy levels of a few
vital molecular species. They may lead to important
observational consequences, which will be discussed in our
following paper.

Compared to the initial conditions of $B_0 = 3~\mu\G$ and
$\beta_0 = 1.23$, the evolved magnetic fields depicted in
Figure~\ref{fig:slice_fiducial} demonstrate a moderate
enhancement due to turbulent dynamics, reaching magnitudes
of approximately $\sim 5 - 10~\mu\G$ predominantly in
regions of compression characterized by relatively high
density ($\rho \gtrsim 10^3~m_p~\cm^{-3}$) and low
temperature ($T \lesssim 20~\K$). Areas where the plasma
beta ($\beta$) is significantly less than unity typically
exhibit magnetic field strengths close to the initial value
of $|B|\sim B_0$. These regions are predominantly composed
of kinematically decelerated gas with densities lower than
the initial density, ($\rho \lesssim 10^{-1}\rho_0$), and
temperatures well above average ($T\sim 10^2~\K$).  The
magnetization within these regions achieves a quasi-steady
state, representing a balance between the amplification
effects of turbulent motion and the counteracting influence
of magnetic diffusion. The lower-right panel presents
$\Rem$, the magnetic Reynolds number for the ambipolar
(``Pederson'') component of magnetic diffusivities, defined
as,
\begin{equation}
  \label{eq:results-Re-m}
  \Rem \equiv \dfrac{|\mathbf{v}| l_{\rm dyn}}{
    \eta_\p } \sim \dfrac{|\mathbf{v}|^2 }{
    \eta_\p |\nabla \cdot \mathbf{v}| }\ ,
\end{equation}
in which we have used the uncapped
$l_v \equiv |\mathbf{v}| / |\nabla \cdot \mathbf{v}|$ to
estimate the dynamical length scale $l_{\rm dyn}$. The
intensity of magnetic diffusion relative to the inertia
becomes more important with smaller $\Rem$. From the figure,
we observe a significant fraction of space where
$\Rem\lesssim 10$ or even less than unity, indicating that
the ambipolar diffusion becomes a major channel of
dissipating the growths of fields through turbulences. These
$\Rem$ values emphasize the necessity of including non-ideal
MHD effects, which can only be accurately modeled by the
inclusion of real-time thermochemistry (see
\S\ref{sec:method-mhd}, \S\ref{sec:method-micro}).

\subsection{Thermochemistry and Its Evolution}
\label{sec:fiducial-chem}

Profiles of chemical species vary in the simulation, from
their abundances to their forming and destruction
pathways. Molecular hydrogen \chem{H_2} are often traced by
other molecules with relatively strong emission powers,
e.g., \chem{CO}, \chem{OH}, and \chem{C^+}. In
Figure~\ref{fig:chem_fiducial}, we present the abundances
relative to hydrogen nuclei $n_\chem{H}$ of a few key
chemical species in the $z=0$ plane for the fiducial model
at $t=5\times 10^4~\yr$, $2\times 10^5~\yr$, and $10^6~\yr$,
respectively.

\subsubsection{Abundances of Molecules}

The relative abundance of $X_\chem{CO}$ for CO, which is the
most frequently adopted \chem{H_2} tracer, varies over time
along the evolution track. Formation of \chem{CO} is lead by
the chain from \chem{C^+} to \chem{CH}
\citep[e.g.][]{DraineBook}, reaching its maximum around
$t\sim 2\times 10^5~\yr$ and starting to decline afterward
(which can also be observed in
Figure~\ref{fig:chem_fiducial}). Such decline comes from the
increased abundance of atomic \chem{H} competes with
\chem{O} in consuming \chem{CH}, which is qualitatively
similar to the tests in Appendix~\ref
{sec:method-static-verify} (see also the evolving \chem{CH}
abundances in Figure~\ref{fig:chem_fiducial}). The timescale
of overall $X_\chem{CO}$ variation is roughly
$\tau_{X_\chem{CO}}\gtrsim 2\times 10^5~\yr$.  It is a
reasonable assumption that diffuse molecular gas coexists
with the overall molecular region, which may subsequently
disperse due to intense photodissociation following the
formation of nearby young stellar objects. The timescale
over which $X_\chem{CO}$ varies is on par with the lifespan
of molecular regions in their entirety; therefore, this
timescale merits close consideration when one intends to
investigate the long-term chemical evolution of diffuse
molecular gases.

\begin{figure*}
  \centering
  \includegraphics[width=7.2in, keepaspectratio]
  {\figdir/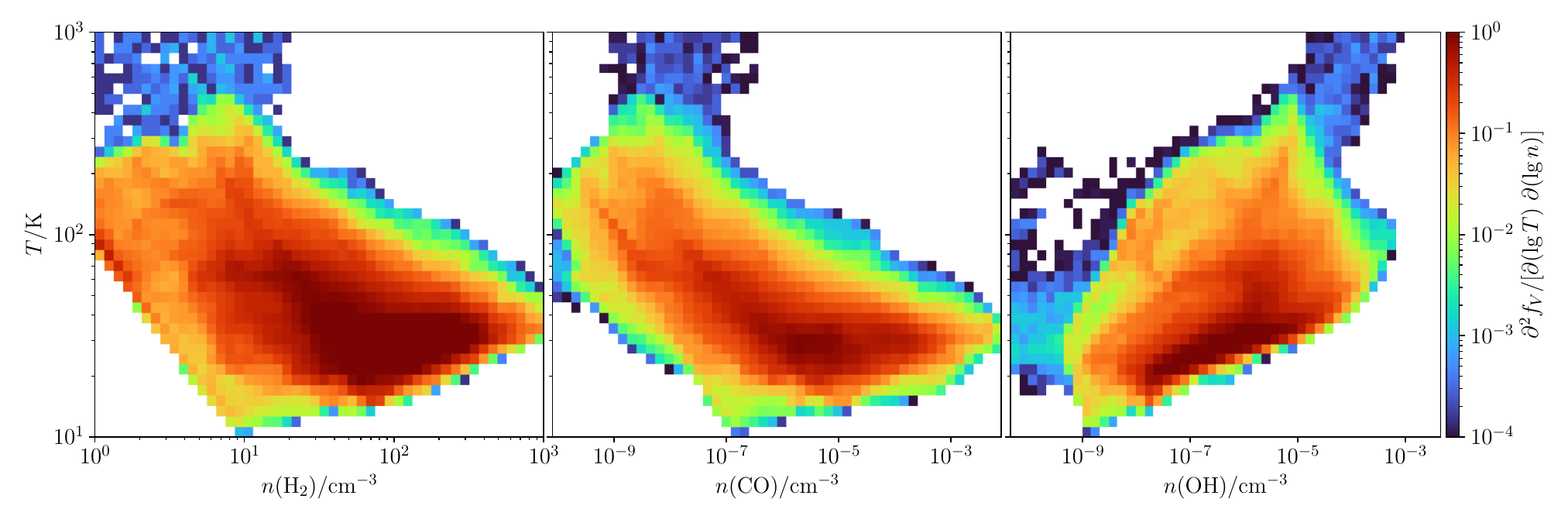}
  \newline
  \includegraphics[width=7.2in, keepaspectratio]
  {\figdir/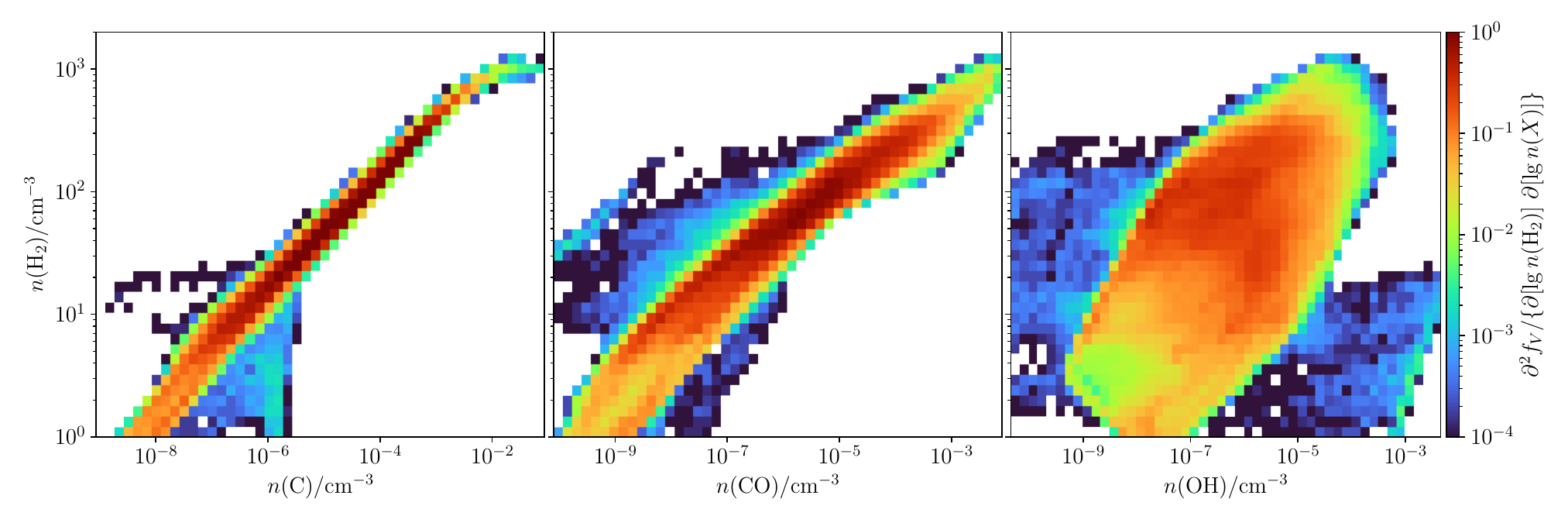}
  \caption{Distribution of spatial volume fraction $f_V$ in
    the spaces spanned by
    $\{\lg T\}\times\{\lg n({\rm X})\}$ (upper row) and
    $\{\lg n(\chem{H_2})\}\times\{\lg n({\rm X})\}$ (lower
    row), where X stands for different chemical species
    indicated on each panel. }
  \label{fig:chem_phase_fiducial}
\end{figure*}

We notice that the \chem{CO} abundance is one order of
magnitude above the hydrostatic tests (Appendix~\ref
{sec:method-static-verify} and
Figure~\ref{fig:uniform-test}). Because the formation of
\chem{CH_2^+}, the rate-determining step in the reaction
chain leading to \chem{CO}, is primarily the two-body
reaction
$\chem{C^+} + \chem{H_2} \rightarrow \chem{CH_2^+}$, the
compression caused by shocks and subsonic motions in
turbulences largely enhances the whole rate of
formation. This enhancement also leads to the spatial
variations in $X_\chem{CO}$, as well as the stochasticity of
$\mean{X_\chem{CO}}$ even though the spatial average is
taken across the domain. In
Figure~\ref{fig:chem_phase_fiducial}, one can observe that
the most of \chem{CO} molecules exist on the average
temperature ($\sim 30~\K$) and relatively high density,
exhibiting distribution contours similar to \chem{H_2} on
the $\lg n-\lg T$ plane. When coming to the
$\lg n-\lg n(\chem{H_2})$ plane, $X_\chem{CO}$ varies by
about one decade at each $n_{\chem{H_2}}$, and the median of
$n_{\chem{CO}}$ increases faster than linear with greater
$n_{\chem{H_2}}$, which is another consequence of the
two-body reaction nature of the rate-determining step.
Therefore, $X_\chem{CO}$ values should be carefully adopted
with their temporal and spatial variations taken into
account adequately when one wants to characterize diffuse
molecular gases with reasonable accuracy. Similar
conclusions about $X_\chem{CO}$ have emerged from various
other simulation studies
\response{\citep[e.g.][]{2019MNRAS.486.4622C,
    2021A&A...654L...6L}} .  These studies emphasize the
integration of co-evolved chemistry with
magnetohydrodynamics (MHD) when analyzing chemical tracers
within the molecular ISM. There are also extra complications
at temperatures below the threshold of $T \lesssim 20~\K$,
CO molecules are prone to further condensation onto the
surfaces of dust grains \citep{2017ApJ...849...80c}. Given
that our study predominantly concentrates on the gas-phase
characteristics of diffuse molecular regions, we have opted
to exclude these additional complexities in the present
paper. Instead, we intend to reserve more exhaustive
explorations for future research endeavors that delve into
multi-phase mechanisms, thereby providing a more
comprehensive understanding of the interplay between gas and
dust in these environments.

The \chem{OH} molecules show contrasts to \chem{CO}. Because
of turbulence compressions, they are $\sim 10^4$ more
abundant than the hydrostatic tests in
Appendix~\ref{sec:method-static-verify}. Nonetheless,
because \chem{e^-} and atomic \chem{O} compete for the
crucial progenitors (\chem{H_2^+} and \chem{H_3^+}), regions
with high densities could not always gain more \chem{OH}
from increased two-body reaction rates. At higher
temperatures, the reaction between \chem{e^-} and
\chem{H_x^+} ($x=2, 3$) is slightly suppressed (rate
coefficient $\propto T^{-0.5}$; see \citealt{umist2013})
while the channel with \chem{O} leading to \chem{H_2O^+} and
then \chem{OH} is not affected. In addition, the charge
exchange channel ($\chem{H^+}+\chem{O} \rightarrow
\chem{O^+} + \chem{H}$) that has secondary importance in the
formation of \chem{OH} prefers higher temperatures and
higher \chem{H^+} abundances, which mostly occur on the
edges of high-density regions where gas compression is
likely to happen. Therefore, the distribution of \chem{OH}
on the $\lg n$-$\lg T$ plane prefers the compressed regions
with intermediately low temperature and densities, showing a
``ridge'', roughly starts from the lower-left part (low $n$,
low $T$) to the upper-right corner
(Figure~\ref{fig:chem_phase_fiducial}). On the $\lg n-\lg
n(\chem{H_2})$ plane, the correlation between \chem{OH} and
\chem{H_2} is loose and should not be used alone to deduce
the properties of \chem{H_2}.

\subsubsection{Energy Dissipation}

\begin{figure*} 
  \centering
  \includegraphics[width=7.2in, keepaspectratio]
  {\figdir/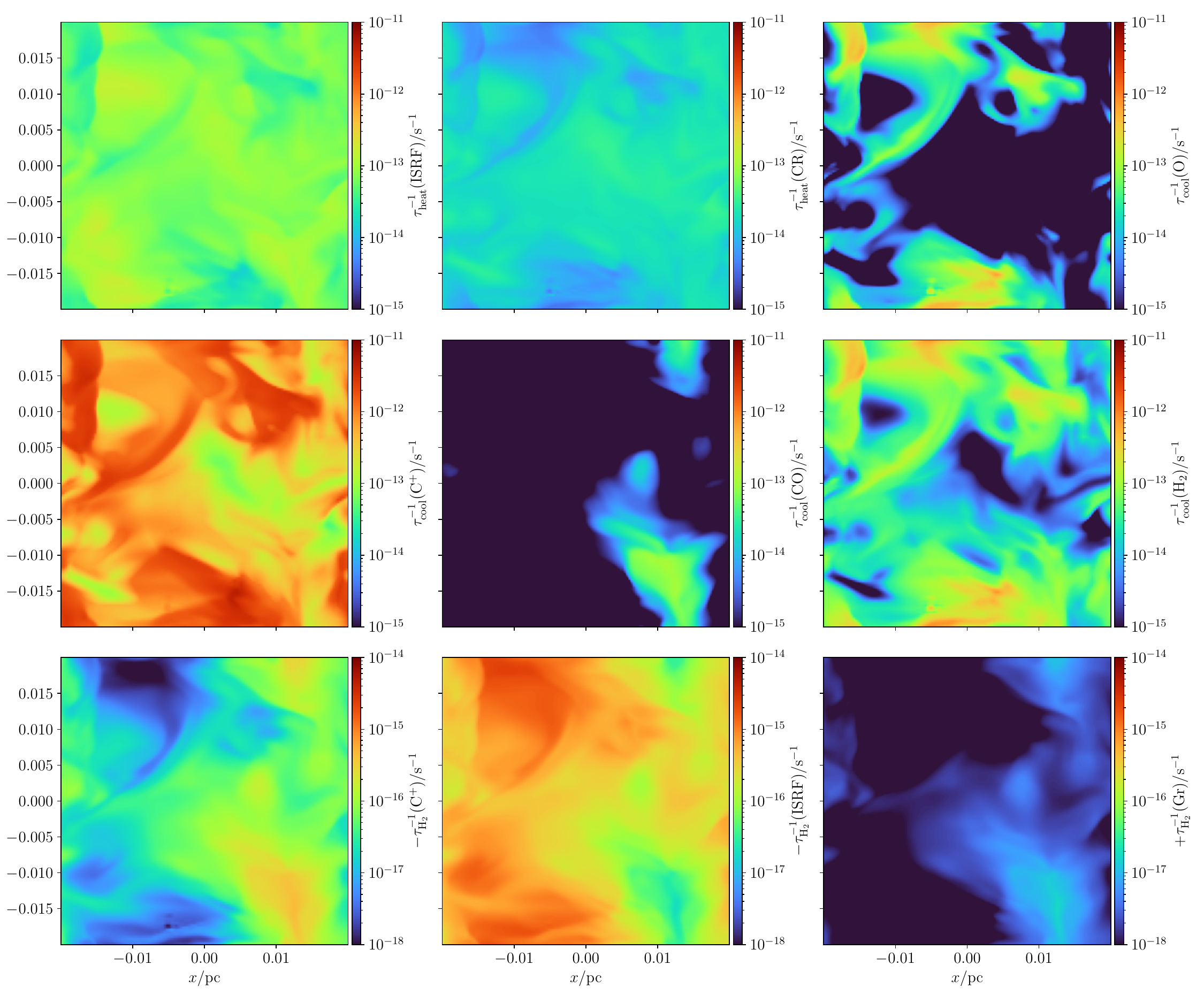}
  \caption{Important thermochemical rates, quantified as the
    reciprocals of timescales ($\tau^{-1}$), illustrating
    the $z=0$ plane for the fiducial model at the
    $t=2\times 10^6~\yr$ snapshot.  {\bf The top and middle
      rows} show the heating and cooling processes. The
    upper-left panel illustrates the heating via the ISRF
    dissociation and ionization, the upper-middel panel
    presents the heating by cosmic rays, and the rest panels
    are for cooling: the fine structures transitions of
    \chem{O} (upper-right panel) and \chem{C^+} (middle-left
    panel), and the ro-vibrational transitions of \chem{CO}
    (middle panel) and \chem{H_2} (middle-right panel). The
    cooling rates by \chem{OH} are relatively tiny and thus
    omitted. {\bf The bottom row} shows the rates related to
    \chem{H_2} number densities: the destruction rates via
    the reactions with \chem{C^+} (left panel) and
    photodissociation (middle panel), and the formation
    rates via the catalyzed reaction by dust grains (right
    panel). }
  \label{fig:thermchem_fiducial}
\end{figure*}

Dissipation of energy is crucial to the quasi-steady
turbulent structures in the system. Short of cooling
mechanisms, the fluids will eventually stop behaving
turbulently after their total internal energy ($\epsilon_g$)
exceeds the amount of energy injection from the largest
scales. Enumerating all prospectively important cooling
processes in the simulation, we identify a few major cooling
mechanisms in the fiducial model and present their
corresponding heating and cooling timescale maps in the top
and middle rows of Figure~\ref{fig:thermchem_fiducial}.

It is observed that cooling via the fine structure
transitions of \chem{C^+} mainly appears at boundaries
separating dense and diffuse regions, while the \chem{CO}
ro-vibrational transitions are responsible for cooling the
dense gas. The rotational transitions of \chem{CO} are
characterized by lower energy separations.  For instance,
the transition $J=1\rightarrow 0$ corresponds to a
wavelength of $\lambda = 2.6~{\rm mm}$. This is in contrast
to the fine structure transition of ionized carbon
(\chem{C^+}), particularly the [\ion{C}{II}] line at
$\lambda = 158 \micron$ via which most of the \chem{C^+}
cooling takes place. The distinction between these two
cooling channels is more apparent at relatively low
temperatures $T\lesssim 30~\K$.  This natural divergence at
cooler regimes has been previously discussed by e.g.
\citet{2014MNRAS.440.3349R, 2014MNRAS.442.2780R}, who
utilized thermochemical calculations in the absence of
hydrodynamics or MHD. In the context of co-evolved non-ideal
MHD, this phenomenon is further intertwined with the
turbulence dynamics. The compression of gas through
turbulence facilitates more efficient production of
\chem{CO} from \chem{C^+} through a sequence of chemical
reactions. Concurrently, the increase in collisional
partners augments the difference in cooling efficiency
between \chem{CO} and \chem{C^+} at low temperatures.

Due to relatively low temperature, the cooling mechanism on
the third panel--the ro-vibrational transitions of
\chem{CO}--are $\sim 10^{-2}$ less efficient than \chem{C^+}
and \chem{H_2}. The detectability of these transitions could
lead to significant observational effects, which will be
discussed in more detail in our following paper. In total,
the evolution of multiple cooling mechanisms is compared to
the energy injection mechanisms: (1) turbulent energy
injection, and (2) the formation of \chem{H_2} on grain
surfaces (each \chem{H_2} formed will introduce $\sim 1~\eV$
into the gas thermal energy; e.g.
\citealt{Pantaleone2021}). The overall energy balance is
close to equilibrium, yet with a small surplus on the
heating side (around $\sim
10^{-4}~\erg~\g~\s^{-1}$). Multiplying this surplus by
$\sim 10^6~\yr$ explains the gradual increase in the total
specific energy (Figure~\ref{fig:history_fiducial} a).

\section{Parameter Studies: Various Models}
\label{sec:var-model}

\begin{deluxetable*}{llccccccccccc}
  \tablecolumns{13}
  \tabletypesize{\scriptsize}
  % \tablewidth{0pt}
  \tablecaption{Physical parameters of
    various models
    (\S\ref{sec:var-model}) \label{table:various_model}}
  \tablehead{ \colhead{} & \colhead{} & \multicolumn{3}{c}
    {$\mean{\epsilon}/(10^9~\erg~\g^{-1})$} & \multicolumn
    {4}{c}{$\log_{10}[n({\rm X})/n_\H]$} &
    \multicolumn{4}{c} {$\log_{10}[\mean{\dot{\epsilon}}_
      {\rm cool}/(\erg~\g^{-1}~\s^{-1})]$}  \\
    \cmidrule(lr){3-5} \cmidrule(lr){6-9}
    \cmidrule(lr){10-13} \colhead{Model} &
    \colhead{Description} & \colhead{Thermal} &
    \colhead{Kinetic} & \colhead{Magnetic} & \colhead{$e^-$}
    & \colhead{\chem{C^+}} & \colhead{\chem{CO}} &
    \colhead{\chem{OH}} & \colhead{\chem{C^+}} &
    \colhead{\chem{CO}} & \colhead{\chem{H_2}} &
    \colhead{\chem{O}} \\
    \colhead{(1)} & \colhead{(2)} & \colhead{(3)} &
    \colhead{(4)} & \colhead{(5)} & \colhead{(6)} &
    \colhead{(7)} & \colhead{(8)} & \colhead{(9)} &
    \colhead{(10)} & \colhead{(11)} & \colhead{(12)} &
    \colhead{(13)} }
  \startdata
  0 & Fiducial model & $2.8$ & $3.0$ & $1.5$ & $-3.99$ &
  $-4.01$ & $-6.58$ & $-7.87$ & $-2.55$ & $-4.73$ & $-3.57$
  & $-4.05$  \\
  \vspace{-0.1cm} \\      
  I & Ideal MHD & $2.7$ & $3.2$ & $1.9$ & $-3.99$ & $-4.01$ &
  $-6.83$ & $-7.84$ & $-2.53$ & $-4.90$ & $-3.68$ & $-4.29$
  \\ 
  B0 & No magnetic fields & $2.7$ & $3.4$ & $0.00$ & $-3.99$
  & $-4.01$ & $-6.61$ & $-7.73$ & $-2.55$ & $-4.67$ &
  $-3.57$ & $-4.13$  \\ 
  B1 & $B_{x0} = 1 \ \mu{\rm G}$ & $2.7$ & $3.4$ & $0.5$ &
  $-4.00$ & $-4.01$ & $-6.33$ & $-7.84$ & $-2.57$ & $-4.46$
  & $-3.52$ & $-4.04$  \\
  B10 & $B_{x0} = 10\ \mu{\rm G}$ & $2.8$ & $1.7$ & $9.9$ &
  $-3.99$ & $-4.01$ & $-6.52$ & $-8.06$ & $-2.54$ & $-4.81$
  & $-3.77$ & $-4.22$  \\
  \vspace{-0.1cm} \\      
  T4 & $\dot{\epsilon}_\turb = 3 \times 10^{-4} {\rm\ erg\
    s}^{-1}{\rm g}^{-1}$ & $1.9$ & $0.6$ & $1.1$ & $-3.98$ &
  $-4.01$ & $-7.05$ & $-9.35$ & $-3.35$ & $-5.53$ & $-4.91$
  & $-6.43$  \\ 
  T2 & $\dot{\epsilon}_\turb = 3 \times 10^{-2} {\rm\ erg\
    s}^{-1}{\rm g}^{-1}$ & $6.5$ & $13.1$ & $2.6$ & $-4.02$
  & $-4.03$ & $-6.45$ & $-7.01$ & $-1.77$ & $-4.03$ &
  $-1.99$ & $-2.59$  \\ 
  \vspace{-0.1cm} \\    
  N4 & $\rho_0 = 2.5\times 10^4\ m_p{\rm\ cm}^{-3}$ & $2.8$
  & $3.3$ & $0.1$ & $-6.21$ & $-6.53$ & $-4.36$ & $-8.21$ &
  $-3.93$ & $-2.60$ & $-3.35$ & $-4.36$  \\ 
  N6 & $\rho_0 = 2.5\times 10^6\ m_p{\rm\ cm}^{-3}$ & $2.4$
  & $3.2$ & $0.01$ & $-8.10$ & $-9.49$ & $-4.40$ & $-10.45$
  & $-6.53$ & $-2.56$ & $-3.59$ & $-5.62$  \\
  \vspace{-0.1cm} \\    
  G0 & $G = 0$ & $3.9$ & $3.2$ & $0.9$ & $-5.36$ & $-5.82$ &
  $-4.56$ & $-6.32$ & $-3.29$ & $-2.97$ & $-3.02$ & $-3.33$
  \\ 
  G3 & $G = 3\ G_0$ & $3.0$ & $2.9$ & $1.5$ & $-3.98$ &
  $-4.00$ & $-8.79$ & $-8.69$ & $-2.45$ & $-6.67$ & $-3.52$
  & $-3.91$  \\
  \vspace{-0.1cm} \\    
  CR0 & $\xi_{\rm CR}=0$ & $2.7$ & $2.9$ & $1.4$ & $-4.02$ &
  $-4.02$ & $-5.94$ & $-10.04$ & $-2.57$ & $-4.32$ & $-3.59$
  & $-4.05$  \\ 
  CR15 & $\xi_{\rm CR}=10^{-15}\ {\rm s}^{-1}$ & $3.8$ &
  $2.9$ & $1.7$ & $-3.53$ & $-4.00$ & $-7.48$ & $-6.74$ &
  $-2.11$ & $-5.32$ & $-3.27$ & $-3.73$  \\
  \vspace{-0.1cm} \\  
  M01 & $0.1\times$ Fiducial metal & $3.7$ & $2.9$ & $1.1$ &
  $-4.65$ & $-5.00$ & $-8.46$ & $-7.61$ & $-2.69$ & $-6.30$
  & $-2.99$ & $-3.86$  \\ 
  M10 & $10 \times$ Fiducial metal & $2.8$ & $3.0$ & $1.6$ &
  $-3.44$ & $-3.44$ & $-5.93$ & $-8.06$ & $-2.55$ & $-4.09$
  & $-3.62$ & $-4.27$  \\ 
  \enddata
  \tablecomments{(1)--(2): Model codes and descriptions;
    each model is different from the fiducial
    model by only one physical parameter. \\
    (3)--(5) Average specific energy components (thermal
    $\mean{\epsilon_g}$, kinetic $\mean{\epsilon_k}$, and
    magnetic $\mean{\epsilon_b}$). Note that for a
    predominantly molecular gas, the specific energy
    components can be converted to temperature as
    $\mean{T} \simeq 11.4~\K~
    [\mean{\epsilon_g}/(10^9~\erg~\g^{-1})]$, and
    root-mean-squared velocity
    $\mean{v^2}^{1/2}\simeq 0.45~\km~\s^{-1}
    [\mean{\epsilon_k}/(10^9~\erg~\g^{-1})]^{1/2}$. The
    average plasma $\beta$ can also be estimated by
    $\mean{\beta} \simeq \mean{\epsilon_g} /
    \mean{\epsilon_b}$.
    \\
    (6)--(9): Logarithms of relative abundances of key
    chemical species.\\
    (10)--(13): Specific cooling rates (energy loss rate per
    unit mass) via different coolants. }
\end{deluxetable*} 

As \S\ref{sec:fiducial} discovered, various physical
parameters are related to the behaviors in the simulated
diffuse cloud. In order to examine their prospective
influences, we conduct extra simulations examining the
outcomes of modifying important physical
parameters. Table~\ref{table:various_model} lists relevant
models marked by different symbols; each varies only one
parameter at a time.

\subsection{Fields and Fluids}
\label{sec:var-model-mag}

\begin{figure}
  \centering
  \includegraphics[width=3.5in, keepaspectratio]
  {\figdir/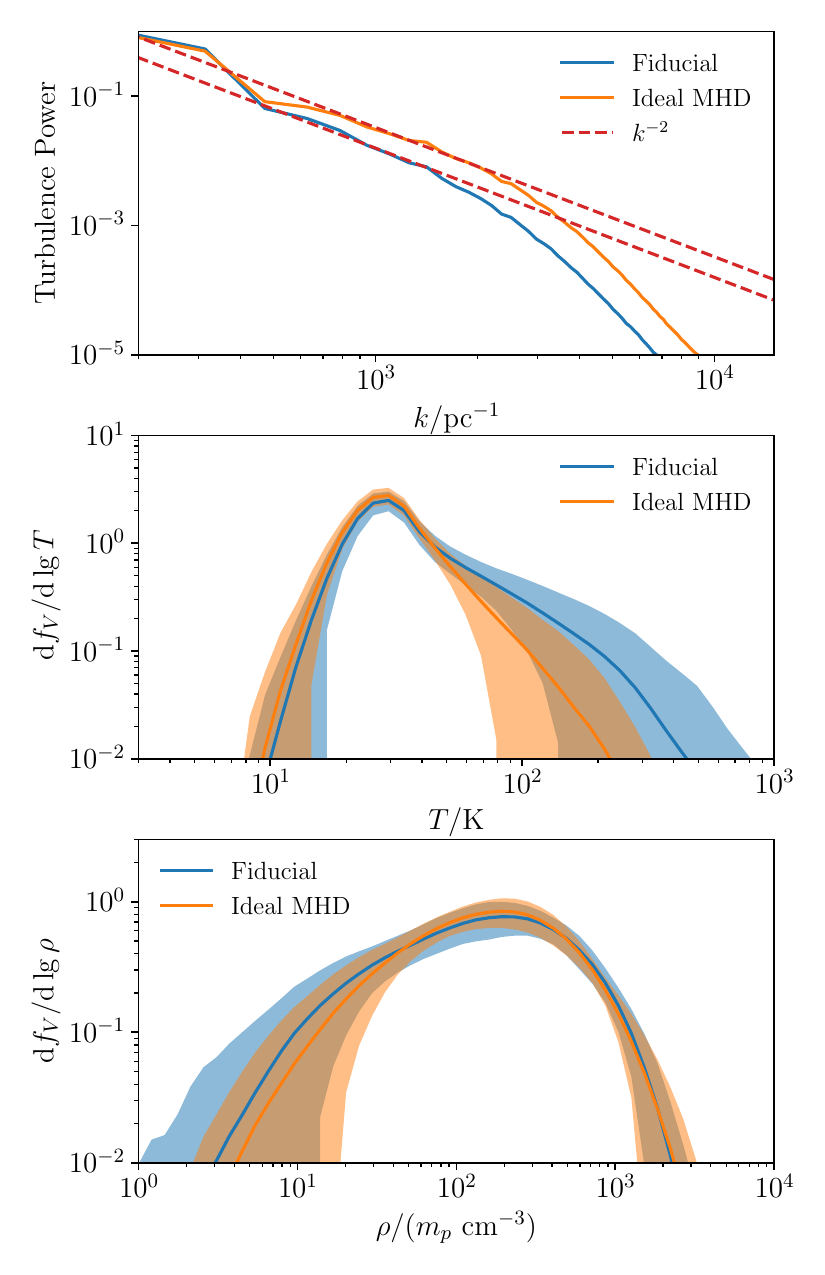}
  \caption{Statistics of the simulation series exploring the
    effect of non-ideal MHD effects by comparing Models 0
    (fiducial) and I (ideal MHD), showing the kinetic energy
    spectra (top panel; an $E(k)\propto k^{-2}$ line is also
    presented for reference), $\lg T$ distributions (middle
    panel), and $\lg\rho$ distributions (lower panel). }
  \label{fig:var_stat_ideal}
\end{figure}

\begin{figure}
  \centering
  \includegraphics[width=3.5in, keepaspectratio]
  {\figdir/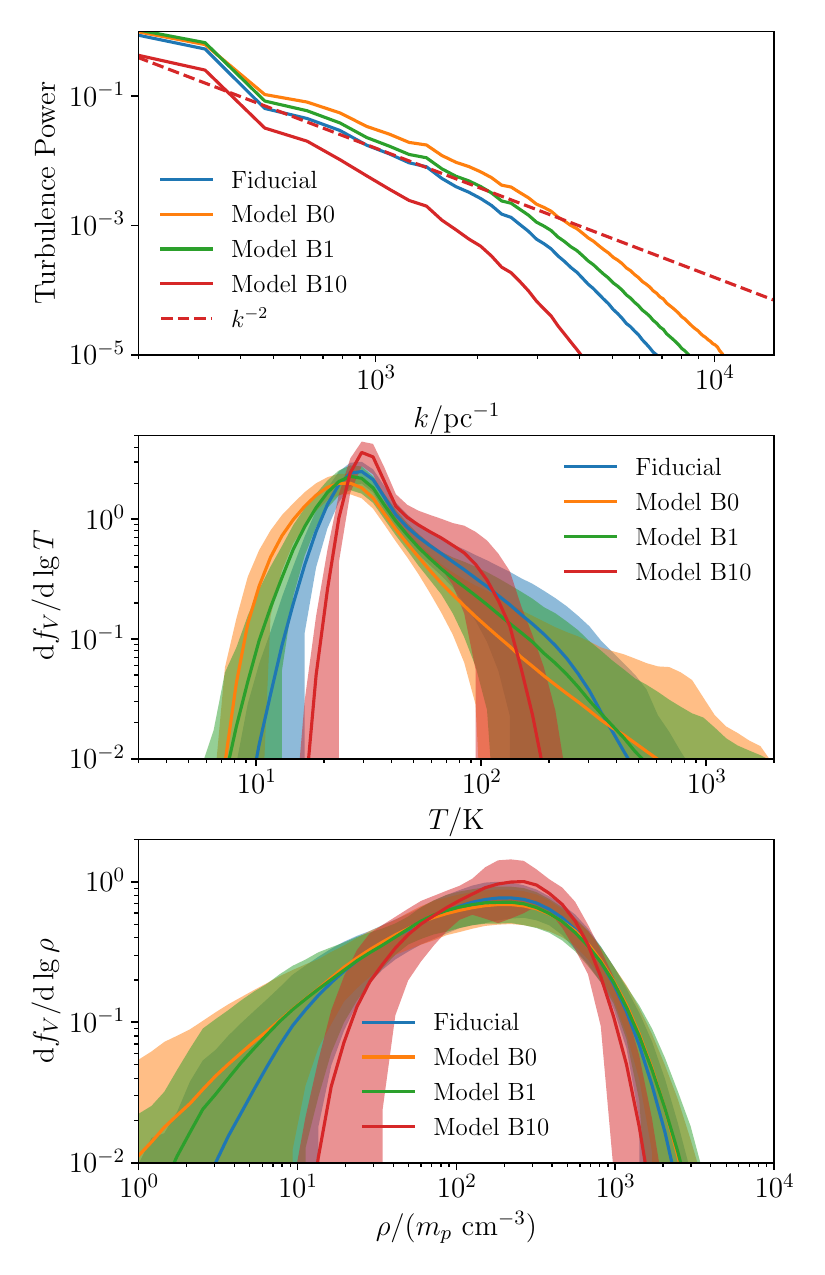}
  \caption{Similar to Figure~\ref{fig:var_stat_ideal} but for
  the series with different initial magnetic field strengths
  (Models 0, B0, B1, and B10). }
  \label{fig:var_stat_field}
\end{figure}

Turbulences can amplify the fields by flux freezing. This
process is dispersed and dissipated by the non-ideal MHD
effects, and we confirm this point by comparing Model 0 and
Model I (ideal MHD). Figure~\ref{fig:var_stat_ideal}
illustrates the difference between these two
models. Although the power spectra are identical at the
spatial scales of energy injection (viz. lowest
wavenumbers), the inertial range of Model I has greater
kinetic power, $\sim 3$ times Model 0. The damping
wavenumber of the fiducial Model 0 is also lower
($k\simeq 1.3\times 10^3~\pc^{-1}$) compared to Model I
($k\simeq 2.2\times 10^3~\pc^{-1}$), exhibiting the effects
of magnetic diffusion that dissipates the kinetic energy at
larger scales with greater efficiencies. The distribution
function in the $\lg T$ space for Model I shows that a
considerably smaller fraction of gas has temperatures above
$\sim 10^2~\K$ compared to Model 0, which is attributed to
the absence of non-ideal MHD heating, as well as the tighter
fluid-field coupling that reduces the shocks and
collisions. These differences confirm the importance of
including proper and consistent non-ideal MHD effects when
simulating magnetized molecular clouds.

Coupling between fluids and fields has another side effect:
the energy transfer to smaller scales is
inhibited. Turbulent motions on smaller scales tend to
amplify the fields more efficiently, causing higher
branching ratios into magnetic energy. In the simulation
series including Models B0 (a pure hydrodynamic simulation
without any fields), fiducial, B1, and B10, the power
spectra are very similar in the range
$k\lesssim 1\times 10^3~\pc^{-1}$. At the same time, the
amplitudes at higher wavenumbers are anti-correlated to the
initial field strengths. The situation of Model B10 is more
severe than others, whose strong magnetic fields suppress
the kinetic power by almost one decade and narrow the
distribution functions in the $\lg T$ and $\lg \rho$ spaces.
This phenomenon can constrain the upper limits of field
intensities in observed molecular gases. For example, the
average field strength is unlikely to be greater than
$\sim 10~\mu\G$ given the discovery of considerable
fractions of $T\gtrsim 1.5\times 10^2~\K$ warm regions.
With the rotational transition of \chem{H_2} observed
identifying gas temperature $T \gtrsim 200~\K$ in the
diffuse molecular gas in the TMC region
\citep[e.g.][]{2010ApJ...715.1370G}, this result
prospectively provide a constraint on the strength of
magnetization therein.

Recent simulations have delved into the intricacies of
magnetized molecular clouds, with notable studies such as
that by \citet{2021ApJ...909..148G}, which explored the
suppression of kinetic energy cascade due to magnetic
tension. This finding aligns well with the observed trends
across a range of simulations. However, it is crucial to
acknowledge that the numerical diffusivities associated with
both fluids and fields, which vary with spatial resolution
in these simulations, may be interlinked with the
identification of damping effects attributed to magnetic
tension.  For a thorough and quantitative examination of how
small-scale structure growth is curtailed by magnetic
fields, it is essential to ensure a consistent coupling
between fields and fluids. This should be governed by
appropriate magnetic diffusivities and should include a
consistent approach to the ultimate dissipation of energy
through thermodynamic processes.  In the current work, the
suppression of energy cascade by field tensions is observed
to influence the power spectra at higher magnetic field
strengths. The threshold for this effect is distinctly
identifiable between magnetic field strengths of
$B_0 = 3~\mu\G$ and $10~\mu\G$. Below this threshold, the
damping of fluid fluctuations predominantly occurs through
non-ideal MHD dissipation mechanisms. This can be inferred
by comparing the damping wavenumbers across Models B3, B10,
0 (Fiducial), and I (ideal MHD).

\begin{figure}
  \centering
  \includegraphics[width=3.5in, keepaspectratio]
  {\figdir/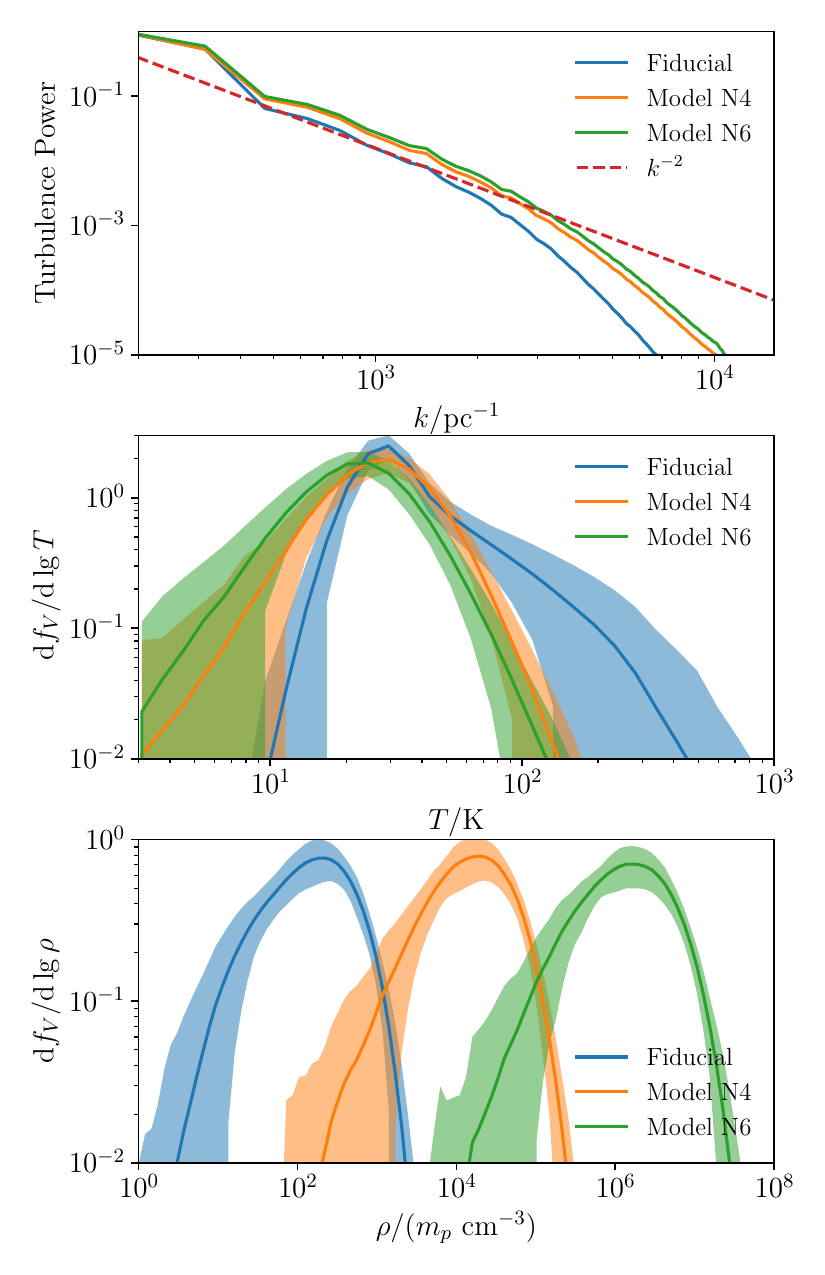}   
  \caption{Similar to Figure~\ref{fig:var_stat_ideal} but
    for the series with different average mass densities
    (Models 0, N4 and N6). }
  \label{fig:var_stat_density}
\end{figure}

Higher average mass density yields large plasma $\beta$,
lowering the relative significance of magnetic pressure and
stresses. As a result, Models N4 and N6 show very similar
power spectra, which have considerably greater amplitudes
compared to the fiducial Model 0 (see
Figure~\ref{fig:var_stat_density})--these spectra are also
very similar to the field-free Model B0. As the cooling
rates are roughly proportional to the square of mass density
at a fixed temperature, while the turbulence energy
injection rate per unit volume is proportional to the mass
density itself (given fixed $\dot{\epsilon}_\turb$ per unit
mass), the temperature distributions of Models N4 and N6
appear on the lower end compared to Model 0.

Another important process affected by the average density is
the recombination of ions. As the recombination rate
coefficients are mostly the functions of temperatures,
assuming that the temperature is roughly fixed and that
there is only one ionizable species
($\chem{X} \leftrightarrow e^- + \chem{X^+}$) thus
$n_e = n(\chem{X^+})$, the ionization--recombination balance
roughly reads,
\begin{equation}
  \label{eq:ion-recomb-bal}
  \alpha x_e^2 n_\H^2 \sim \zeta n_\H
  \Rightarrow x_e\sim n_\H^{-1/2},\
  n_e \equiv x_e n_\H \sim n_\H^{1/2} .
\end{equation}
In the series consisting of Models 0, N4, and N6, the
fraction $x_e$ drops faster than $\rho^{-1/2}$. The main
charge carrier in Model N4 before $t\sim 2\times 10^5~\yr$
is still \chem{C+}, and changes to \chem{H^+} afterward due
to efficient formation of \chem{CO} that uses up almost all
carbon. Model N6 also has \chem{H^+} as the main charge
carrier, but from a different origin: its self-shielding and
cross-shielding for the ISRF ionization of neutral \chem{C}
is significant that the dominating ionization process is the
cosmic ray ionization of \chem{H_2} in the first
place. Without efficient \chem{CO} formation, the specific
energy keeps growing until $\sim 10^5~\yr$, after which
$\mean{X_\chem{CO}}$ exceeds $\sim 10^{-5}$, and the
subsequent cooling causes a continuous drop in gas
temperature over the following $10^6~\yr$.

\subsection{Turbulence Energy Injection}
\label{sec:var-model-turb}

Turbulent energy injection (see \S\ref{sec:method-turb}) is
the ultimate energy source in all simulations throughout
this paper. By turning the knob that controls the injection
rate $\dot{\epsilon}_\turb$ in Models T2 and T4 (injection
rates $3\times 10^{-2}$ and $3\times 10^{-4}$ in units
$\erg~\g^{-1}~\s^{-1}$, respectively), the turbulence
cascade and almost all phenomena attached to it will
experience considerable changes constrained by their energy
budgets.

\begin{figure}
  \centering
  \includegraphics[width=3.5in, keepaspectratio]
  {\figdir/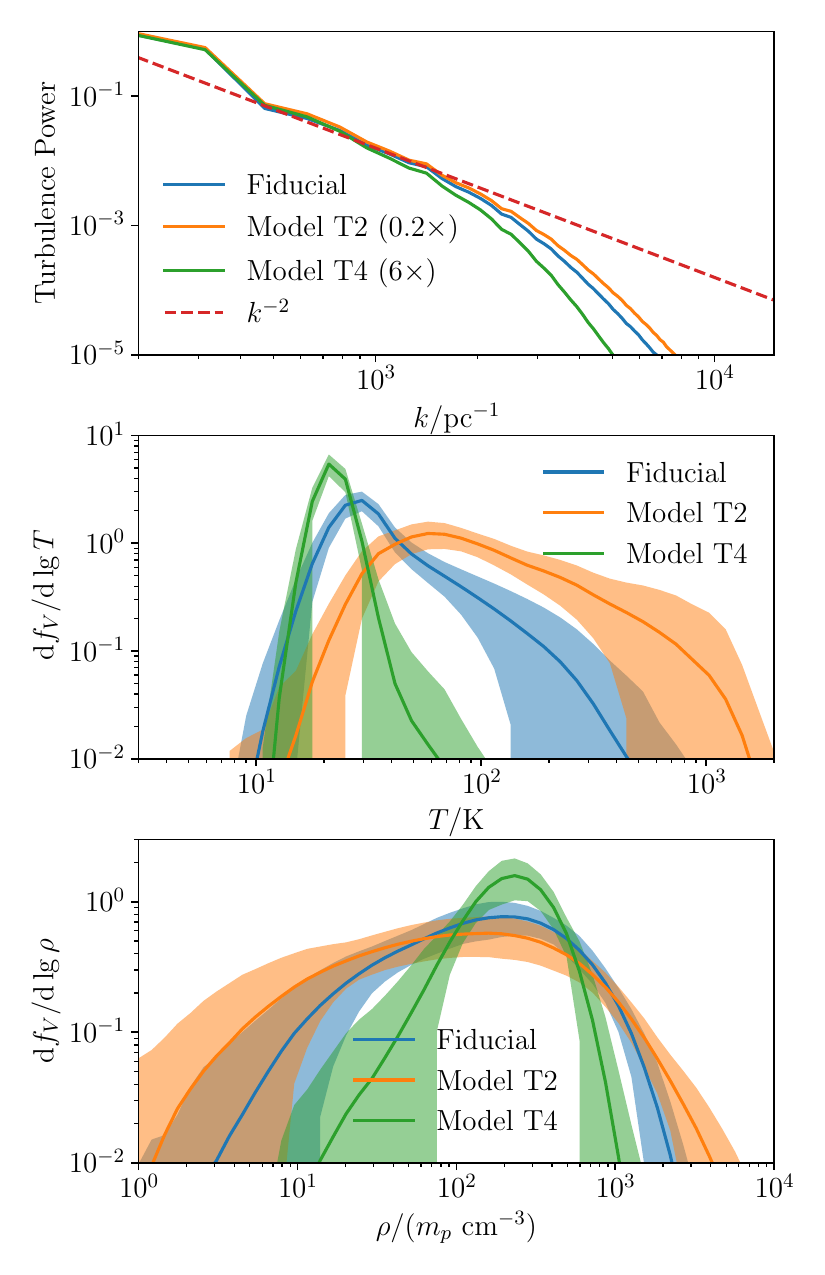}
  \caption{Similar to Figure~\ref{fig:var_stat_ideal} but
    for the series with different turbulence energy
    injection rates (Models 0, T2, and T4). \response{Note
      that the power spectra for T2 and T4 have adjusted by
      scaling factors (indicated in the legend), so that
      their spectra on the lowest wavenumbers overlap for
      better comparison.} }
  \label{fig:var_stat_turb}
\end{figure}

The most prominent consequence of $\epsilon_\turb$ is the
distribution of gas dynamics and
thermodynamics. Figure~\ref{fig:var_stat_turb} compares
Models T2 and T4 with the fiducial Model 0 for their
statistical functions. \response{The shapes of the kinetic
  energy spectra are almost identical at wavenumbers below
  the damping wavenumber $k\simeq 1.5\times 10^3~\pc^{-1}$,
  while the amplitude grows sublinearly with increasing
  intensity of turbulence energy cascades. Above that wave
  number, the cases with stronger energy injection are
  relatively less susceptible to the inhibition of the
  kinetic power by the damping mechanisms including magnetic
  fields and diffusivities on smaller spatial scales, and
  vice versa.}

% similar throughout this series, while the Model T2 with
% higher $\dot{\epsilon}_\turb$ has a higher cutoff
% wavenumber at $k\sim 3\times 10^3~\pc^{-1}$
% ($l\sim 5\times 10^{-4}~\pc$) compared to the fiducial
% model's cutoff $k\sim 1.5\times 10^3~\pc^{-1}$ and Model
% T4's $k\sim 10^3~\pc^{-1}$. Stronger kinetic energy
% injection tends to overwhelm the magnetic inhibition of
% turbulence structures at smaller spatial scales (see also
% the discussions in \S\ref{sec:var-model-mag}).

The density and temperature distributions in this series
exhibit a clear trend that higher $\dot{\epsilon}_\turb$
leads to ``broader'' distributions in both $\lg T$ and
$\lg \rho$ space by allowing for more vigorous shocks,
compressions, and adiabatic expansions. This fact suggests a
promising way to quantify the turbulence energy crossing
rates via the measurements of temperature and density
distributions of concerned molecular gases. For instance,
the temperature distributions are reasonably sensitive to
$\dot{\epsilon}_\turb$, as Model T4 exhibits little gas
warmer than $\sim 60~\K$, while Model T2 has a
non-negligible fraction with $T\gtrsim 10^3~\K$. Admittedly,
multiple other factors affect gas temperature distributions,
causing parametric degeneracies. Therefore, quantitative
measurement of the turbulence power relies on other
constraints, e.g., the abundance of \chem{OH} is positively
correlated to $\dot{\epsilon}_\turb$
(Table\ref{table:various_model}), due to the enhanced
reaction rates at higher density and temperatures. Synthetic
mock observations are necessary in the following works to
confront the simulation results directly with observations.

\subsection{Thermochemical Conditions}
\label{sec:var-chem}

\begin{figure}
  \centering
  \includegraphics[width=3.5in, keepaspectratio]
  {\figdir/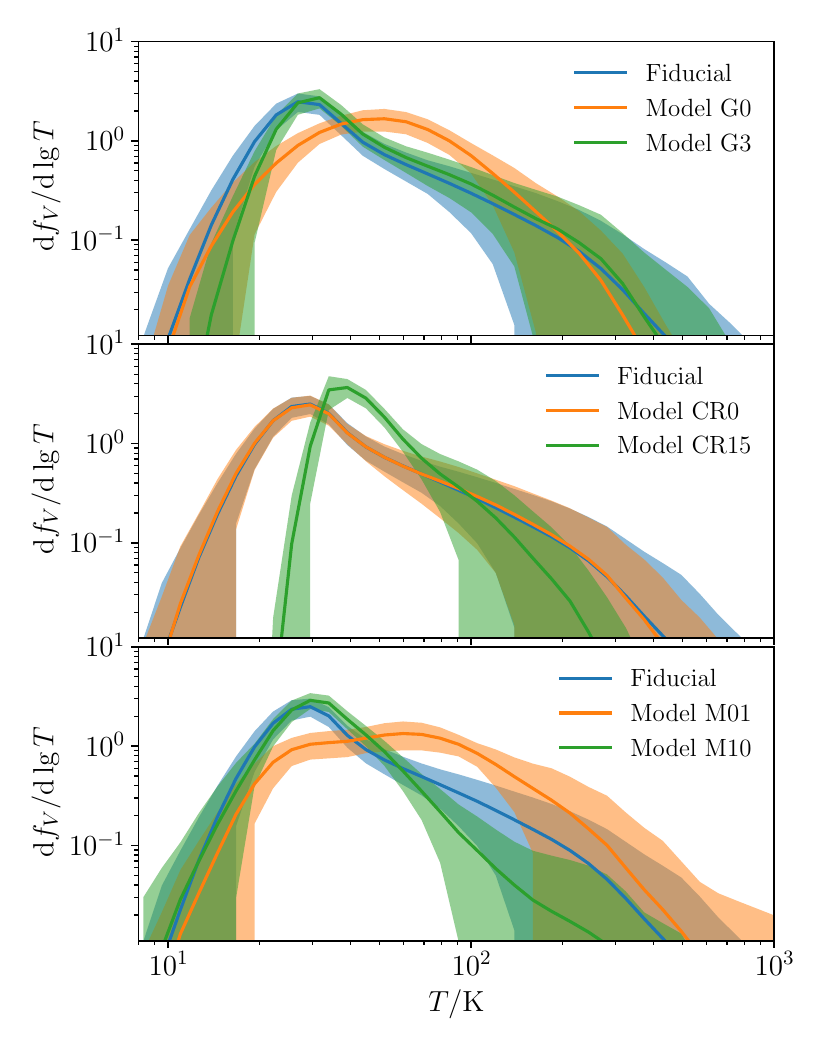}    
  \caption{Similar to the middle panel of
    Figure~\ref{fig:var_stat_ideal}, with models varying the
    cosmic ray intensities (left column; Models 0, CR0, and
    CR15), and the ISRF (right column; Models 0, G0, and
    G3). The power spectra and $\lg \rho$ distributions are
    omitted as they are very similar to the fiducial model.}
  \label{fig:hist_lgT_multi}
\end{figure}

There are two major ionization channels in the simulation:
(a) ISRF ionization of carbon and (b) cosmic ray ionization
(mainly on \chem{H_2}, \chem{H} and \chem{He}). These
processes not only provide necessary coupling between gas
and fields but also effectively initiate the formation of
relevant chemical species, including \chem{C^+},  \chem{CO},
\chem{OH}, which are the most important ones in terms of the
gas thermodynamics.

The statistics of simulations have revealed that the kinetic
energy spectra and the $\lg \rho$ distribution functions of
these models are affected only very slightly. Therefore,
Figure~\ref{fig:hist_lgT_multi} only emphasizes the
comparisons in the $\lg T$ space. Model G3 has a relatively
similar $\lg T$ distribution function compared to Model 0,
with the lower end slightly warmer due to the increased ISRF
destruction of molecular coolants (\chem{CO} and
\chem{OH}). Model G0, with the lack of production in the
most critical coolants \chem{C^+} and \chem{CO}, exhibits a
$\lg T$ profile that mostly fits the standard log-normal
distribution. With excessive cosmic ray ionization, Model
CR15 has a $\sim 2.5\times$ higher cutoff on the
low-temperature end due to extra cosmic ray heating, while
its high end of $\lg T$ distribution is qualitatively
similar to Model I due to higher abundances of charge
carriers and tighter fluid-field couplings. Since the most
important coolants are often metal-related, altering the
gas's metallicity impacts the gas temperatures. Models 0,
M01, and M10 have almost the same lower limit in the
temperature distribution functions (approximately
$T\sim 10~\K$), while their high ends are anti-correlated
with the metallicity. In brief, the thermodynamic conditions
can provide specific constraints on the properties of the
irradiation and cooling processes in molecular gases.

\section{Discussions and Summary}
\label{sec:summary}

In this work, we have studied the characteristics and
evolution of diffuse ($\rho\sim 250~m_p~\cm^{-3}$),
magnetized ($B_0 = 3~\mu G$) molecular clouds, using MHD
simulations co-evolved with consistent non-equilibrium
thermochemistry. Although mostly numerical and theoretical,
these simulations have revealed prospective clues that can
play crucial roles in constraining diffuse molecular gas
properties via observational measurements.

\subsection{Fields, Fluids, and their Interactions}
\label{sec:summary-dynamics}

Co-evolved with thermochemistry, we observe from the
simulation models that the behaviors of gas dynamics are
predominantly controlled by the turbulence and MHD
properties themselves. The $\lg T$ distributions (especially
their high-temperature cutoff) are hopefully the most
sensitive indicator of the turbulence energy injection
rate. With consistent thermochemistry considered, the
thermal energy budgets at quasi-steady states are mostly
balanced, and the temperature profiles will reveal the
dynamics relatively robustly. The $\lg T$ profiles are
susceptible to other factors, including total densities,
ionizing radiations (and cosmic rays), field intensities,
and metallicities. Nevertheless, as discussed in
\S\ref{sec:var-model}, these factors only affect $\lg T$
with secondary importance, especially regarding the
high-temperature cutoffs in these diffuse molecular
gases. In observations aiming at constraining the gas
dynamic properties, various nebula diagnostics can determine
the gas temperature distribution in interested regions and
constrain the turbulence energy injections with relatively
high confidence.

The field strengths we have studied are marginally crucial
in curbing the formation of turbulence structures, except
for Model B10, whose average plasma $\beta\lesssim 10^{-1}$
causes significant inhibition of turbulent structure
growths. The simulated models are mostly weakly ionized, and
non-ideal MHD is desired to model the kinematics and
dynamics. Due to magnetic diffusions, hydrodynamic profiles
are not very sensitive probes in the fields. However, the
upper limit of fields can still be constrained by the
temperature assisted with other measurements.

\subsection{The Necessity of Non-ideal MHD Studies with
  Consistent Thermochemistry}

\subsubsection{Magnetic Fields }
  
There have been many recent numerical simulations aiming at
the turbulence dynamics within the interstellar medium
(ISM), including the behavior within molecular clouds. A
prime example is the work by \citet{2021NatAs...5..365F},
which achieved remarkable resolution through hydrodynamic
simulations, albeit without incorporating magnetic
fields. One of the major conclusions of this study is the
crucial role that magnetic fields play in sculpting the
turbulence power spectra, especially the damping
wavenumber. The magnetic pressure and tension are found to
impede the development of structures at smaller scales.  In
scenarios devoid of magnetic fields, the energy cascade is
observed to extend to the highest attainable wavenumbers,
constrained only by numerical dissipations at the resolution
limits. However, this overlooks the actual spatial scales at
which damping occurs, given that magnetization is a
pervasive characteristic of the ISM under
consideration. This oversight underscores the necessity for
simulations to account for the influence of magnetic fields
to more accurately reflect the ISM's behavior.

\subsubsection{Magnetic Diffusivities}

Even the simulations incorporating MHD, such as
\citet{1999ApJ...513..259O, 2015ApJ...810...93P,
  2019PhRvL.122m5101B, 2021ApJ...909..148G}, often overlook
a critical aspect: the influence of non-ideal MHD
effects. As depicted in Figure~\ref{fig:slice_fiducial} and
corroborated by numerous other simulations, regions that are
susceptible to non-ideal MHD effects (indicated by $\Rem$ at
the order of unity) constitute a significant portion of the
simulated volume, particularly at the peripheries of
compressed gas clumps. This phenomenon specifically impacts
the morphology at smaller scales, as illustrated by the
comparison between Model 0 and Model I in
Figure~\ref{fig:var_stat_ideal}. The damping scale,
influenced by magnetic diffusivities, is approximately twice
as extensive as in scenarios modeled with ideal MHD, where
the effective magnetic diffusivities are numerical artifacts
contingent on resolution.

Also elaborated by \citet{2023ApJ...942L..34G}, the
dissipation scales in MHD play a pivotal role in determining
how magnetic fields sculpt the dynamics of the gas. The
accurate incorporation of magnetic diffusivities in
simulations that resolve the spatial scales of diffusion is
essential to circumvent such artifacts. Moreover, magnetic
diffusivities are sensitively dependent on the actual
thermochemical conditions of the gas, including the
concentration and types of charge carriers, the densities of
neutral gas, and the prevailing temperatures. These physical
variables are contingent on the evolutionary history of the
gases and should not be reduced to mere status variables.
Consequently, non-ideal MHD simulations of molecular gases
that employ prescribed parametrization of diffusivities or
highly simplified chemical networks, as seen in
\citet{2009ApJ...701.1258D, 2011MNRAS.415.3681T,
  2011MNRAS.418..390J, 2012MNRAS.420..817J,
  2016MNRAS.457.1037W}, inevitably sacrifice
consistency. This is due to the fact that many reactions
related to ionization, recombination, and charge transfer
involve complex molecular and ionic species, and sometimes
even surface chemical processes. This realization has
already been acknowledged in studies of protoplanetary disks
\citep[e.g.,][]{bai+goodman2009, 2016ApJ...819...68X,
  2019ApJ...874...90W} and has been identified as being
equally significant for the diffuse molecular gases
discussed in this paper, as exemplified in
Figure~\ref{fig:hist_lgT_multi}.

\subsubsection{Equation of State and Thermodynamics}

The majority of prior research works, either with
hydrodynamics or MHD, have employed an isothermal equation
of state (EoS), neglecting the essential dissipation
mechanisms carried out by comprehensive thermochemical
approaches. It is acknowledged that simulations relying
solely on an adiabatic EoS in the absence of cooling will
eventually convert all injected energy into thermal energy
of the gas, overwhelming the kinetic and magnetic energy and
finally smearing out the turbulent structures.  However, the
use of an isothermal EoS as a simplification also
significantly compromises physical fidelity.

Consider the scenario of a gas clump that undergoes rapid
compression prior to cooling. The gas pressure, and
consequently its internal energy density, scales with the
length scale $l$ of the clump as $p\proptosim l^{-3\gamma}$,
while the density scales as $\rho \proptosim l^{-3}$. The
cooling rate per unit volume at a constant temperature is
proportional to $\Lambda \rho^2\proptosim l^{-6}$, a steeper
power-law than that of the gas pressure.  This indicates
that the compressed gas patch will undergo runaway cooling
until it reaches a significantly lower temperature, at which
$\Lambda$ almost diminishes compared to its ambient value
(for molecular gas, this temperature is typically around
$\sim 10~\K$ ), while other regions of the gas remain at
higher temperatures.  This phenomenon results in a broad
temperature distribution that spans several orders of
magnitude, as observable in e.g.,
Figures~\ref{fig:slice_fiducial} and
Section~\ref{sec:var-model}.

The presence of a broad temperature distribution within
turbulent materials is essential for interpreting the
diagnostics of observed nebulae. For instance, the work by
\citet{2010ApJ...715.1370G} has already identified the
emission of rotational transitions of \chem{H_2} from
diffuse molecular regions with a median temperature
$\lesssim 50~\K$. This finding underscores the importance of
regions heated to temperatures above $\gtrsim 10^2~\K$
through gas dynamics, thereby highlighting the need for
consistent thermodynamics in ISM simulations.  In contrast,
the simplified dichotomy of ``supersonic versus subsonic
scales'' as emphasized in isothermal simulations, including
those by \citet{2021NatAs...5..365F}, does not accurately
reflect the complexities of astrophysical phenomena taking
place in the diffuse molecular gas.

When aligning simulation results with ISM observations, the
acquisition of consistent temperature profiles is imperative
for the precise interpretation of spectral diagnostics
pertaining to the properties of the gas. These
considerations underscore the necessity for the integration
of robust thermochemical computations within simulation
models, moving beyond the limitations of the overly
simplified isothermal equation of state. This approach is
vital for capturing the nuanced interplay of physical
processes that govern the behavior of the interstellar
medium.

\subsubsection{Dimensionless versus Dimensional}
  
Most of the simulations on turbulent ISM follow the
dimensionless approach.  When the final dissipation is based
on realistic microphysical mechanisms (including magnetic
diffusion, radiative cooling, etc.) rather than mathematical
or numerical dissipations, one can easily confirm with basic
dimensional analyses that the microphysical processes cannot
be non-dimensionalized simultaneously with the hydrodynamics
or MHD. This fact inevitably add to the complications and
computational costs, yet this is the inevitable costs that
one has to pay to understand turbulent gases that are
coupled with physical mechanisms that are crucial to the
ultimate dissipation along the turbulence energy cascade,
including magnetic fields and multi-scale physics.

\subsection{Thermochemical Probes of Diffuse Molecular
  Gases}
\label{sec:summary-chem}

% We already observed a significant increase in

Given relatively strong turbulence driving processes,
constraining thermochemistry via dynamic profiles are
indirect and may suffer from degeneracies of
parameters---probably except for the influence by
metallicities on the higher limits of temperatures, which
can be constrained in conjunction with the specific
turbulence powers via the upper limits in $T$. Instead,
thermochemistry itself is the probe of multiple dynamical
processes. This probe can only be forged with turbulent ISM
simulations that can directly yield analyses consistently
with real-time thermochemistry.  Adopting isothermal EoS may
mislead the interactions connecting the gas dynamics to
chemistry and subsequent observables, by providing
inconsistent key parameters including compression ratios and
temperatures.

One prospectively interesting indicator that deserves extra
effort is the \chem{CO} relative abundances,
$X_{\chem{CO}}$. Presumably, this is mostly a constant used
to infer total gas densities in many studies, yet our
simulations suggest that $X_{\chem{CO}}$ varies from place
to place, from time to time. The conversion from \chem{C^+}
to \chem{CO} takes $\sim 2\times 10^5~\yr$ to equilibrate
the photodissociation. Before that, the reduced \chem{CO}
relative abundance could be a helpful indicator limiting the
evolution history of the gas. This constraint should be
confirmed with the measurement of \chem{C^+} and \chem{CH},
which are essential materials and intermediate products in
the reaction chains for \chem{CO} formation.

Another promising probe is \chem{OH}, which is a sharp
indication of cosmic ionization of \chem{H_2} (and atomic
\chem{O} as well, with roughly half of the importance in the
chemical network). Due to similar ionization energy values
for hydrogen and oxygen, efficient ionization can only come
from high-penetration sources, including the cosmic ray or
the external X-ray irradiation from, e.g., young protostars
(not included in this paper). Since the relative abundance
of \chem{OH} keeps climbing throughout the
$\sim 2\times 10^5~\yr$ period at the beginning of
evolution, the measurements could also help constrain the
age of the system. The reaction \chem{H_2O^+} + \chem{H_2}
$\rightarrow$ \chem{H_3O^+} + \chem{H}, one of the slowest
steps in the \chem{OH} formation chain in such weakly
ionized media, is insensitive to temperature. In the
meantime, the photodissociation of \chem{OH} is another
limiting factor, as the process relies on the strong
electric dipole and is not susceptible to self or
cross-shielding. Both factors make the final product
\chem{OH} a possible indication of a few physical
parameters, including local density and the ISRF
intensity. Consistent interpretations of observations will
require simulations for reasonable fittings on the
measurement results.

\subsection{Future Works}

Our current work focuses on numerical simulations, their
intrinsic MHD processes, and the properties based on the
interactions between MHD, radiation (ISRF and cosmic ray),
and thermochemistry. In future works, we will elaborate on
possible predictions by synthesizing mock observations. One
possibility is the line intensities, either spatially
resolved or integrated. Combined with consistent MHD and
thermodynamics, the spatial distribution features of
physical diffuse molecular cloud models can be converted
into expectations on actual astronomical
measurements. Another possibility is the distribution
profiles in the position-position and
position-position-velocity (PPV) spaces. By confronting
observables with synthetic profiles generated from an
adequate number of simulations, plausible interpretations of
measurements will be achieved via searching the parameter
space, supported by proper microphysical details that
interact consistently with the dynamics and their
observational tracers.

\bigskip

\noindent N. Yue, L. Wang, T. Bisbas, and D. Quan are funded
by Zhejiang Laboratory Scientific Project \#
K2022PE0AB01. We also acknowledge the computing resources
provided by the Kavli Institute for Astronomy and
Astrophysics in Peking University. We thank our collegues
for helpful discussions: Paul Goldsmith, Xin Wang, and
Haifeng Yang.

\bibliography{mol_cloud_sim}
\bibliographystyle{aasjournal}

\appendix

\section{Hydrostatic Models}
\label{sec:method-static-verify}

\begin{figure}
  \centering
  \hspace*{-0.4cm}
  \includegraphics[width=3.5in, keepaspectratio]
  {\figdir/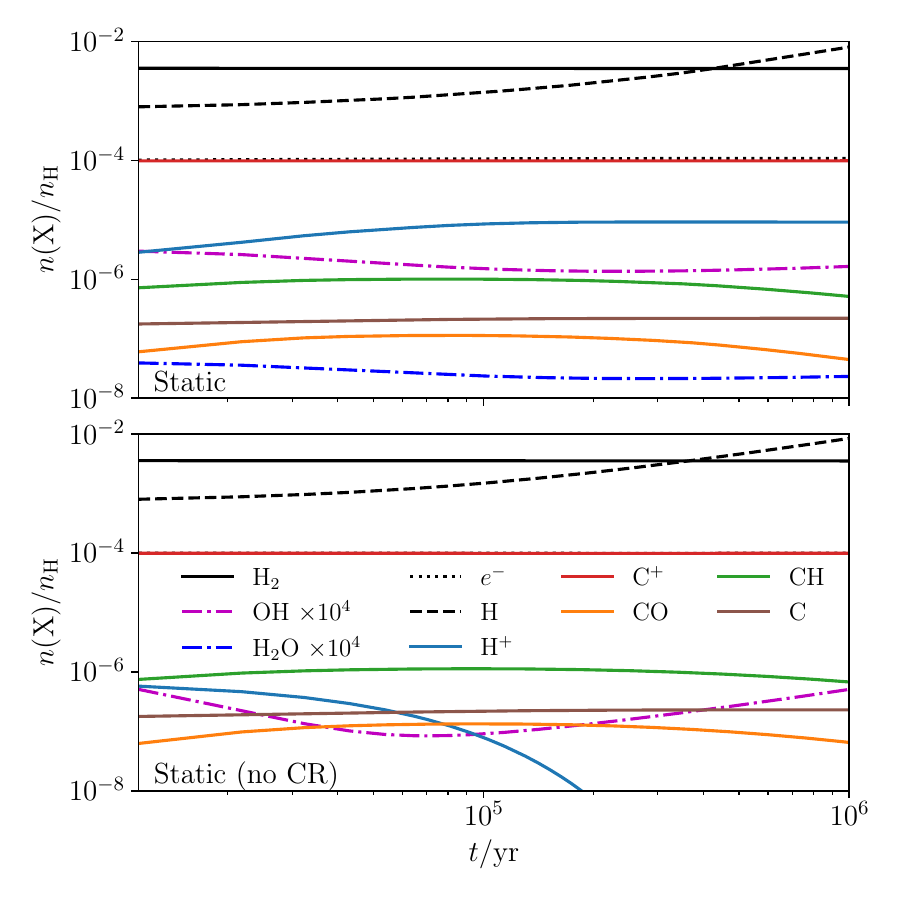} 
  \caption{Similar to the Panel (b) of
    Figure~\ref{fig:history_fiducial} but for the two test
    simulation models under hydrostatics (viz. no
    turbulences or any kinematics). The two panels present
    the results with and without cosmic rays, respectively.}
  \label{fig:uniform-test}
\end{figure}

Figure~\ref{fig:uniform-test} presents the two tests' most
abundant and essential chemical species. These test runs are
identical to the fiducial model except that they are static
(i.e., have {\it no} turbulence injection), and one of them
has zero cosmic ray intensity as well (the lower panel). In
order to verify that our simulation system---already
verified in various non-ideal MHD simulations for PPDs and
planetary atmospheres---is also functioning correctly for
the diffuse molecular clouds, this appendix elaborates on
the consistency of a few key species by comparing the
simulated key values of chemical kinetics with analytic
results.

We first examine the species related to hydrogen. Both
models exhibit the dissociation of hydrogen molecules over
the $10^6~\yr$ evolution, at the end of which the abundance
of atomic \chem{H} reaches $\sim 10^{-2}$ due to the
photodissociation by the ISRF. As the self-shielding of
\chem{H_2} is considered with the effective shielding length
$l_{\rm cap}\sim 1~\pc$ on a uniform medium (see also
\S\ref{sec:method-micro}) with $\rho = 250~m_p~\cm^{-3}$ (or
$n(\chem{H_2})\simeq 200~\cm^{-3}$) and $T=30~\K$, one gets
the self-shielding factor
$f_{\chem{H_2}}\simeq 2.5\times 10^{-5}$ using the recipes
in \citet{1996ApJ...468..269D}. This leads to a
$\zeta_{\rm diss,\chem{H_2}} \simeq 3\times
10^{-16}~\s^{-1}$ photodissociation rate, yielding
$n(\chem{H})\simeq 1.9~\cm^{-3}\simeq 10^{-2}~n_\H$.  The
free electrons are predominantly provided by the ISRF
ionization of \chem{C}, whose abundance almost equals the
elemental abundance of carbon. The \chem{H^+} concentration
is eliminated by the radiative recombination (whose rate
coefficient is
$\alpha_\H \simeq 2\times 10^{-11}~\cm^3~\s^{-1}$ at
$30~\K$) at the timescale
$\tau_\chem{H^+} = (n_e\alpha_\H)^{-1} \simeq 4.6\times
10^4~\yr$, in the absence of cosmic rays. With
$\zeta_{\rm CR} = 10^{-17}~\s^{-1}$, radiative recombination
maintains
$n(\chem{H^+}) \simeq n_\H \zeta_{\rm CR} \tau_\chem{H^+}
\simeq 1.2\times 10^{-3}~\cm^{-3}$. These results agree
quantitatively with the test models.

The formation of \chem{CO} follows a chain of reactions from
\chem{C^+} to \chem{CH} (see e.g., \citealt{DraineBook}),
starting from
$\chem{C^+} + \chem{H_2} \rightarrow \chem{CH_2^+}$. At
$30~\K$ and the \ref{table:fiducial-model} abundances, this
reaction has a $\sim 4.6\times 10^{-15}~\cm^{-3}~\s^{-1}$
production rate of \chem{CH_2^+}, a quarter of which is
instantly converted to \chem{CH} by dissociative
recombination with $e^-$ (equivalent to a \chem{CH}
production rate of
$\sim 1.2\times 10^{-15}~\cm^{-3}~\s^{-1}$). The final step
$\chem{CH} + \chem{O} \rightarrow \chem{CO} + \chem{H}$
competes with
$\chem{CH} + \chem{H} \rightarrow \chem{C} +
\chem{H_2}$. The sum of these two reactions consume
\chem{CH} at the timescales
$\tau_\chem{CH} \sim 1.3\times 10^{11}~\s$, constraining the
\chem{CH} abundance to be
$n(\chem{CH})\simeq 1.6\times 10^{-4}~\cm^{-3}$, and leading
to a production rate of \chem{CO} of
$\sim 6\times 10^{-16}~\cm^{-3}~\s^{-1}$. Consider the ISRF
photodissociation rate
$\zeta_\chem{CO}\simeq 4\times 10^{-11}~\s^{-1}$ (with
$0.3~G_0$ and the total shielding factor $\sim 0.57$
estimated using the recipes in
\citealt{2017A&A...602A.105H}), such \chem{CO} production
rate leads to
$n(\chem{CO})\simeq 1.5\times 10^{-5}~\cm^{-3}$.  The slight
drop of \chem{CO} abundance since $t\sim 10^5~\yr$ is
attributed to the increase of atomic \chem{H}. In addition
to confirming species abundances with
Figure~\ref{fig:uniform-test}, we also verify that the
analytic estimates of reaction rates are quantitatively
consistent with the outputs given by the test simulations
(not shown in the figures).

% The photodissociation of \chem{CO} yields the neutral
% carbon together with
% $\chem{CH} + \chem{H} \rightarrow \chem{C} + \chem{H_2}$,
% $\chem{CH_2^+} + \chem{e^-} \rightarrow \chem{C} +
% 2\chem{H}$,
% $\chem{C^+} + \chem{e^-} \rightarrow \chem{C}$. 
% at a rate $\sim 1\times 10^{-14}~\cm^{-3}~\s^{-1}$.  Using the
% cross-shielding factor ($\sim 0.88$) estimated using the
% recipes in \citet{2017a&a...602a.105h}, the $0.3~G_0$ ISRF
% destroys neutral carbon at $\zeta_{\chem{C}} \simeq
% 6\times 10^{-11}~\s^{-1}$, and constraints the neutral
% carbon abundance to be 

\end{document}

%% file: preamble.tex
\usepackage{float}
\usepackage{amsmath,amssymb}
\usepackage{hyperref} % For internal hyperlinks
\usepackage{graphicx} % For included graphics
\usepackage{color}
\usepackage{verbatim}    
\usepackage{enumitem}
\usepackage{natbib}
\usepackage{multirow}
\newcommand{\proptosim}{\mathrel{\vcenter{
 \offinterlineskip\halign{\hfil$##$\cr
 \propto\cr\noalign{\kern2pt}\sim\cr\noalign{\kern-2pt}}}}}
\renewcommand{\b}[1]{\boldsymbol{#1}}

\newcommand{\response}[1]{{#1}}
\newcommand{\mean}[1]{\langle #1\rangle}

\renewcommand{\min}{\mathrm{min}}

\newcommand{\cm}{\mathrm{cm}}
\newcommand{\m}{\mathrm{m}}

\newcommand{\g}{\mathrm{g}}
\newcommand{\G}{\mathrm{G}}
\newcommand{\K}{\mathrm{K}}  
\newcommand{\km}{\mathrm{km}}

\renewcommand{\micron}{\mathrm{\mu m}}

\newcommand{\pc}{\mathrm{pc}}

\newcommand{\erg}{\mathrm{erg}}
\newcommand{\eV}{\mathrm{eV}}

\newcommand{\s}{\mathrm{s}}
\newcommand{\yr}{\mathrm{yr}}

\renewcommand{\d}{\mathrm{d}}
\renewcommand{\m}{\mathrm{m}}
\newcommand{\f}{\mathrm{f}}
\newcommand{\e}{\mathrm{e}}

\newcommand{\Gr}{\ensuremath{\mathrm{Gr}}}
\renewcommand{\ion}[2]
{{\rm#1}\;\textsc{\MakeLowercase{#2}}}

\newcommand{\p}{{\rm p}}

\newcommand*\chem[1]{\ensuremath{\mathrm{#1}}}
\renewcommand{\neg}[1]{\ensuremath{\mathrm{#1}^-}}

%% file: mol_cloud_sim.bbl
\begin{thebibliography}{}
\expandafter\ifx\csname natexlab\endcsname\relax\def\natexlab#1{#1}\fi
\providecommand{\url}[1]{\href{#1}{#1}}
\providecommand{\dodoi}[1]{doi:~\href{http://doi.org/#1}{\nolinkurl{#1}}}
\providecommand{\doeprint}[1]{\href{http://ascl.net/#1}{\nolinkurl{http://ascl.net/#1}}}
\providecommand{\doarXiv}[1]{\href{https://arxiv.org/abs/#1}{\nolinkurl{https://arxiv.org/abs/#1}}}

\bibitem[{{Bai}(2011)}]{2011ApJ...739...51B}
{Bai}, X.-N. 2011, \apj, 739, 51, \dodoi{10.1088/0004-637X/739/1/51}

\bibitem[{{Bai} \& {Goodman}(2009)}]{bai+goodman2009}
{Bai}, X.-N., \& {Goodman}, J. 2009, \apj, 701, 737

\bibitem[{{Bian} \& {Aluie}(2019)}]{2019PhRvL.122m5101B}
{Bian}, X., \& {Aluie}, H. 2019, \prl, 122, 135101,
  \dodoi{10.1103/PhysRevLett.122.135101}

\bibitem[{{Boldyrev} \& {Perez}(2009)}]{2009PhRvL.103v5001B}
{Boldyrev}, S., \& {Perez}, J.~C. 2009, \prl, 103, 225001,
  \dodoi{10.1103/PhysRevLett.103.225001}

\bibitem[{{Cazaux} {et~al.}(2017){Cazaux}, {Mart{\'\i}n-Dom{\'e}nech}, {Chen},
  {Mu{\~n}oz Caro}, \& {Gonz{\'a}lez D{\'\i}az}}]{2017ApJ...849...80c}
{Cazaux}, S., {Mart{\'\i}n-Dom{\'e}nech}, R., {Chen}, Y.~J., {Mu{\~n}oz Caro},
  G.~M., \& {Gonz{\'a}lez D{\'\i}az}, C. 2017, \apj, 849, 80,
  \dodoi{10.3847/1538-4357/aa8b0c}

\bibitem[{{Clark} {et~al.}(2019){Clark}, {Glover}, {Ragan}, \&
  {Duarte-Cabral}}]{2019MNRAS.486.4622C}
{Clark}, P.~C., {Glover}, S. C.~O., {Ragan}, S.~E., \& {Duarte-Cabral}, A.
  2019, \mnras, 486, 4622, \dodoi{10.1093/mnras/stz1119}

\bibitem[{{Downes} \& {O'Sullivan}(2009)}]{2009ApJ...701.1258D}
{Downes}, T.~P., \& {O'Sullivan}, S. 2009, \apj, 701, 1258,
  \dodoi{10.1088/0004-637X/701/2/1258}

\bibitem[{{Draine}(2011)}]{DraineBook}
{Draine}, B.~T. 2011, {Physics of the Interstellar and Intergalactic Medium}
  (Princeton University Press)

\bibitem[{{Draine} \& {Bertoldi}(1996)}]{1996ApJ...468..269D}
{Draine}, B.~T., \& {Bertoldi}, F. 1996, \apj, 468, 269, \dodoi{10.1086/177689}

\bibitem[{{Federrath} {et~al.}(2021){Federrath}, {Klessen}, {Iapichino}, \&
  {Beattie}}]{2021NatAs...5..365F}
{Federrath}, C., {Klessen}, R.~S., {Iapichino}, L., \& {Beattie}, J.~R. 2021,
  Nature Astronomy, 5, 365, \dodoi{10.1038/s41550-020-01282-z}

\bibitem[{{Franeck} {et~al.}(2018){Franeck}, {Walch}, {Seifried}, {Clarke},
  {Ossenkopf-Okada}, {Glover}, {Klessen}, {Girichidis}, {Naab}, {W{\"u}nsch},
  {Clark}, {Pellegrini}, \& {Peters}}]{2018MNRAS.481.4277F}
{Franeck}, A., {Walch}, S., {Seifried}, D., {et~al.} 2018, \mnras, 481, 4277,
  \dodoi{10.1093/mnras/sty2507}

\bibitem[{{Glover} \& {Jappsen}(2007)}]{2007ApJ...666....1G}
{Glover}, S.~C.~O., \& {Jappsen}, A.~K. 2007, \apj, 666, 1,
  \dodoi{10.1086/519445}

\bibitem[{{Goldsmith} {et~al.}(2010){Goldsmith}, {Velusamy}, {Li}, \&
  {Langer}}]{2010ApJ...715.1370G}
{Goldsmith}, P.~F., {Velusamy}, T., {Li}, D., \& {Langer}, W.~D. 2010, \apj,
  715, 1370, \dodoi{10.1088/0004-637X/715/2/1370}

\bibitem[{{Grete} {et~al.}(2021){Grete}, {O'Shea}, \&
  {Beckwith}}]{2021ApJ...909..148G}
{Grete}, P., {O'Shea}, B.~W., \& {Beckwith}, K. 2021, \apj, 909, 148,
  \dodoi{10.3847/1538-4357/abdd22}

\bibitem[{{Grete} {et~al.}(2023){Grete}, {O'Shea}, \&
  {Beckwith}}]{2023ApJ...942L..34G}
---. 2023, \apjl, 942, L34, \dodoi{10.3847/2041-8213/acaea7}

\bibitem[{{Heays} {et~al.}(2017){Heays}, {Bosman}, \& {van
  Dishoeck}}]{2017A&A...602A.105H}
{Heays}, A.~N., {Bosman}, A.~D., \& {van Dishoeck}, E.~F. 2017, \aap, 602,
  A105, \dodoi{10.1051/0004-6361/201628742}

\bibitem[{{Hennebelle} \& {Falgarone}(2012)}]{2012A&ARv..20...55H}
{Hennebelle}, P., \& {Falgarone}, E. 2012, \aapr, 20, 55,
  \dodoi{10.1007/s00159-012-0055-y}

\bibitem[{{Ingalls} {et~al.}(2011){Ingalls}, {Bania}, {Boulanger}, {Draine},
  {Falgarone}, \& {Hily-Blant}}]{2011ApJ...743..174I}
{Ingalls}, J.~G., {Bania}, T.~M., {Boulanger}, F., {et~al.} 2011, \apj, 743,
  174, \dodoi{10.1088/0004-637X/743/2/174}

\bibitem[{{Jappsen} {et~al.}(2007){Jappsen}, {Glover}, {Klessen}, \& {Mac
  Low}}]{2007ApJ...660.1332J}
{Jappsen}, A.~K., {Glover}, S.~C.~O., {Klessen}, R.~S., \& {Mac Low}, M.~M.
  2007, \apj, 660, 1332, \dodoi{10.1086/513085}

\bibitem[{{Jones} \& {Downes}(2011)}]{2011MNRAS.418..390J}
{Jones}, A.~C., \& {Downes}, T.~P. 2011, \mnras, 418, 390,
  \dodoi{10.1111/j.1365-2966.2011.19491.x}

\bibitem[{{Jones} \& {Downes}(2012)}]{2012MNRAS.420..817J}
---. 2012, \mnras, 420, 817, \dodoi{10.1111/j.1365-2966.2011.20095.x}

\bibitem[{{Kim} {et~al.}(2018){Kim}, {Kim}, \&
  {Ostriker}}]{2018ApJ...859...68K}
{Kim}, J.-G., {Kim}, W.-T., \& {Ostriker}, E.~C. 2018, \apj, 859, 68,
  \dodoi{10.3847/1538-4357/aabe27}

\bibitem[{{Kim} {et~al.}(2019){Kim}, {Kim}, \&
  {Ostriker}}]{2019ApJ...883..102K}
---. 2019, \apj, 883, 102, \dodoi{10.3847/1538-4357/ab3d3d}

\bibitem[{{Kowal} {et~al.}(2007){Kowal}, {Lazarian}, \&
  {Beresnyak}}]{2007ApJ...658..423K}
{Kowal}, G., {Lazarian}, A., \& {Beresnyak}, A. 2007, \apj, 658, 423,
  \dodoi{10.1086/511515}

\bibitem[{{Langer} {et~al.}(2010){Langer}, {Velusamy}, {Pineda}, {Goldsmith},
  {Li}, \& {Yorke}}]{2010A&A...521L..17L}
{Langer}, W.~D., {Velusamy}, T., {Pineda}, J.~L., {et~al.} 2010, \aap, 521,
  L17, \dodoi{10.1051/0004-6361/201015088}

\bibitem[{{Lupi} {et~al.}(2021){Lupi}, {Bovino}, \&
  {Grassi}}]{2021A&A...654L...6L}
{Lupi}, A., {Bovino}, S., \& {Grassi}, T. 2021, \aap, 654, L6,
  \dodoi{10.1051/0004-6361/202142145}

\bibitem[{{Mac Low}(1999)}]{1999ApJ...524..169m}
{Mac Low}, M.-M. 1999, \apj, 524, 169, \dodoi{10.1086/307784}

\bibitem[{{Mac Low} {et~al.}(1998){Mac Low}, {Klessen}, {Burkert}, \&
  {Smith}}]{1998PhRvL..80.2754M}
{Mac Low}, M.-M., {Klessen}, R.~S., {Burkert}, A., \& {Smith}, M.~D. 1998,
  \prl, 80, 2754, \dodoi{10.1103/PhysRevLett.80.2754}

\bibitem[{{McElroy} {et~al.}(2013){McElroy}, {Walsh}, {Markwick}, {Cordiner},
  {Smith}, \& {Millar}}]{umist2013}
{McElroy}, D., {Walsh}, C., {Markwick}, A.~J., {et~al.} 2013, \aap, 550, A36

\bibitem[{{McKee} \& {Ostriker}(2007)}]{2007ARA&A..45..565M}
{McKee}, C.~F., \& {Ostriker}, E.~C. 2007, \araa, 45, 565,
  \dodoi{10.1146/annurev.astro.45.051806.110602}

\bibitem[{{Nehm{\'e}} {et~al.}(2008){Nehm{\'e}}, {Gry}, {Boulanger}, {Le
  Bourlot}, {Pineau Des For{\^e}ts}, \& {Falgarone}}]{2008A&A...483..471N}
{Nehm{\'e}}, C., {Gry}, C., {Boulanger}, F., {et~al.} 2008, \aap, 483, 471,
  \dodoi{10.1051/0004-6361:20078373}

\bibitem[{{Neufeld} \& {Kaufman}(1993)}]{1993ApJ...418..263N}
{Neufeld}, D.~A., \& {Kaufman}, M.~J. 1993, \apj, 418, 263,
  \dodoi{10.1086/173388}

\bibitem[{{Neufeld} {et~al.}(1995){Neufeld}, {Lepp}, \&
  {Melnick}}]{1995ApJS..100..132N}
{Neufeld}, D.~A., {Lepp}, S., \& {Melnick}, G.~J. 1995, \apjs, 100, 132,
  \dodoi{10.1086/192211}

\bibitem[{{Omukai} {et~al.}(2010){Omukai}, {Hosokawa}, \&
  {Yoshida}}]{2010ApJ...722.1793O}
{Omukai}, K., {Hosokawa}, T., \& {Yoshida}, N. 2010, \apj, 722, 1793,
  \dodoi{10.1088/0004-637X/722/2/1793}

\bibitem[{{Ostriker} {et~al.}(1999){Ostriker}, {Gammie}, \&
  {Stone}}]{1999ApJ...513..259O}
{Ostriker}, E.~C., {Gammie}, C.~F., \& {Stone}, J.~M. 1999, \apj, 513, 259,
  \dodoi{10.1086/306842}

\bibitem[{{Pantaleone} {et~al.}(2021){Pantaleone}, {Enrique-Romero},
  {Ceccarelli}, {Ferrero}, {Balucani}, {Rimola}, \&
  {Ugliengo}}]{Pantaleone2021}
{Pantaleone}, S., {Enrique-Romero}, J., {Ceccarelli}, C., {et~al.} 2021, \apj,
  917, 49, \dodoi{10.3847/1538-4357/ac0142}

\bibitem[{{Pineda} {et~al.}(2010){Pineda}, {Velusamy}, {Langer}, {Goldsmith},
  {Li}, \& {Yorke}}]{2010A&A...521L..19P}
{Pineda}, J.~L., {Velusamy}, T., {Langer}, W.~D., {et~al.} 2010, \aap, 521,
  L19, \dodoi{10.1051/0004-6361/201015089}

\bibitem[{{Porter} {et~al.}(2015){Porter}, {Jones}, \&
  {Ryu}}]{2015ApJ...810...93P}
{Porter}, D.~H., {Jones}, T.~W., \& {Ryu}, D. 2015, \apj, 810, 93,
  \dodoi{10.1088/0004-637X/810/2/93}

\bibitem[{{Qian} {et~al.}(2018){Qian}, {Li}, {Gao}, {Xu}, \&
  {Pan}}]{2018ApJ...864..116Q}
{Qian}, L., {Li}, D., {Gao}, Y., {Xu}, H., \& {Pan}, Z. 2018, \apj, 864, 116,
  \dodoi{10.3847/1538-4357/aad780}

\bibitem[{{Raskutti} {et~al.}(2016){Raskutti}, {Ostriker}, \&
  {Skinner}}]{2016ApJ...829..130R}
{Raskutti}, S., {Ostriker}, E.~C., \& {Skinner}, M.~A. 2016, \apj, 829, 130,
  \dodoi{10.3847/0004-637X/829/2/130}

\bibitem[{{Raskutti} {et~al.}(2017){Raskutti}, {Ostriker}, \&
  {Skinner}}]{2017ApJ...850..112R}
---. 2017, \apj, 850, 112, \dodoi{10.3847/1538-4357/aa965e}

\bibitem[{{Rice} {et~al.}(2018){Rice}, {Federman}, {Flagey}, {Goldsmith},
  {Langer}, {Pineda}, \& {Lambert}}]{2018ApJ...858..111R}
{Rice}, J.~S., {Federman}, S.~R., {Flagey}, N., {et~al.} 2018, \apj, 858, 111,
  \dodoi{10.3847/1538-4357/aabae7}

\bibitem[{{Richings} {et~al.}(2014{\natexlab{a}}){Richings}, {Schaye}, \&
  {Oppenheimer}}]{2014MNRAS.440.3349R}
{Richings}, A.~J., {Schaye}, J., \& {Oppenheimer}, B.~D. 2014{\natexlab{a}},
  \mnras, 440, 3349, \dodoi{10.1093/mnras/stu525}

\bibitem[{{Richings} {et~al.}(2014{\natexlab{b}}){Richings}, {Schaye}, \&
  {Oppenheimer}}]{2014MNRAS.442.2780R}
---. 2014{\natexlab{b}}, \mnras, 442, 2780, \dodoi{10.1093/mnras/stu1046}

\bibitem[{{Stone} {et~al.}(2020){Stone}, {Tomida}, {White}, \&
  {Felker}}]{2020ApJS..249....4S}
{Stone}, J.~M., {Tomida}, K., {White}, C.~J., \& {Felker}, K.~G. 2020, \apjs,
  249, 4, \dodoi{10.3847/1538-4365/ab929b}

\bibitem[{{Tilley} \& {Balsara}(2011)}]{2011MNRAS.415.3681T}
{Tilley}, D.~A., \& {Balsara}, D.~S. 2011, \mnras, 415, 3681,
  \dodoi{10.1111/j.1365-2966.2011.18982.x}

\bibitem[{{Wang} {et~al.}(2019){Wang}, {Bai}, \&
  {Goodman}}]{2019ApJ...874...90W}
{Wang}, L., {Bai}, X.-N., \& {Goodman}, J. 2019, \apj, 874, 90,
  \dodoi{10.3847/1538-4357/ab06fd}

\bibitem[{{Wang} \& {Dai}(2021)}]{2021ApJ...914...98W}
{Wang}, L., \& {Dai}, F. 2021, \apj, 914, 98, \dodoi{10.3847/1538-4357/abf1ee}

\bibitem[{{Wang} \& {Goodman}(2017)}]{2017ApJ...847...11w}
{Wang}, L., \& {Goodman}, J. 2017, \apj, 847, 11.
\newblock \url{http://stacks.iop.org/0004-637X/847/i=1/a=11}

\bibitem[{{Whitworth} \& {Jaffa}(2018)}]{2018A&A...611A..20W}
{Whitworth}, A.~P., \& {Jaffa}, S.~E. 2018, \aap, 611, A20,
  \dodoi{10.1051/0004-6361/201731871}

\bibitem[{{Wurster} {et~al.}(2019){Wurster}, {Bate}, \&
  {Price}}]{2019MNRAS.489.1719W}
{Wurster}, J., {Bate}, M.~R., \& {Price}, D.~J. 2019, \mnras, 489, 1719,
  \dodoi{10.1093/mnras/stz2215}

\bibitem[{{Wurster} {et~al.}(2016){Wurster}, {Price}, \&
  {Bate}}]{2016MNRAS.457.1037W}
{Wurster}, J., {Price}, D.~J., \& {Bate}, M.~R. 2016, \mnras, 457, 1037,
  \dodoi{10.1093/mnras/stw013}

\bibitem[{{Wurster} \& {Rowan}(2024)}]{2024MNRAS.528.2257W}
{Wurster}, J., \& {Rowan}, C. 2024, \mnras, 528, 2257,
  \dodoi{10.1093/mnras/stae090}

\bibitem[{{Xu} \& {Bai}(2016)}]{2016ApJ...819...68X}
{Xu}, R., \& {Bai}, X.-N. 2016, \apj, 819, 68,
  \dodoi{10.3847/0004-637X/819/1/68}

\end{thebibliography}
